\documentclass[sn-basic]{sn-jnl} 

\usepackage[utf8]{inputenc} 
\usepackage[T1]{fontenc}    
\usepackage{url}            
\usepackage{booktabs}       
\usepackage{amsfonts}       
\usepackage{microtype}      
\usepackage{float}          
\usepackage{multicol}       
\usepackage{multirow}       
\usepackage{amsmath}        
\usepackage{colortbl}       
\usepackage{pgfplots}       
\usepackage{tikz}           
\usepackage[outline]{contour} 
\usepackage{tikz-3dplot}
\usepackage{pgfplotstable}  

\tikzset{>=latex}

\usetikzlibrary{positioning}
\usetikzlibrary{shapes.geometric, decorations.pathreplacing} 
\usetikzlibrary{3d} 
\usetikzlibrary{shadows.blur}
\usetikzlibrary{fadings}
\usetikzlibrary{decorations.text} 
\usetikzlibrary{calc,arrows.meta,positioning}
\usetikzlibrary{matrix}
\definecolor{Grey}{gray}{0.9}
\tikzstyle{xlab}=[below=-1,scale=0.85]
\tikzstyle{ylab}=[left=-1,scale=0.85]

\colorlet{mydarkblue}{blue!40!black}
\colorlet{mylightblue}{mydarkblue!12} 
\colorlet{myred}{red!80!black}
\colorlet{mydarkred}{red!50!black}
\colorlet{mylightred}{mydarkred!12}
\colorlet{mydarkgreen}{green!30!black}
\colorlet{mylightgreen}{mydarkgreen!12}
\colorlet{myorange}{orange!63!black}
\colorlet{mylightorange}{orange!80!black!12}

\def\minwidth{5cm}

\tikzstyle{pooling} = [draw, rectangle, minimum width=\minwidth, minimum height=.75cm, fill=red!20, text centered, anchor=north]

\tikzstyle{activation} = [draw, rectangle, minimum width=\minwidth, minimum height=.75cm, fill=gray!20, text centered, anchor=north]

\tikzstyle{container} = [draw, rectangle, minimum width=\minwidth, minimum height=.75cm, fill=yellow!20, text centered, anchor=north]

\tikzstyle{norm} = [draw, rectangle, minimum width=\minwidth, minimum height=.75cm, fill=blue!20, text centered, anchor=north]

\tikzstyle{conv} = [draw, rectangle, minimum width=\minwidth, minimum height=.75cm, fill=purple!20, text centered, anchor=north]

\tikzstyle{mul} = [draw, rectangle, minimum width=\minwidth, minimum height=.75cm, fill=green!20, text centered, anchor=north]

\tikzstyle{downsample} = [draw, trapezium, trapezium left angle=120, trapezium right angle=120, minimum width = \minwidth, minimum height = .5cm, fill = orange!20, text centered, anchor =north]

\tikzstyle{upsample} = [draw, trapezium, trapezium left angle= 60, trapezium right angle=60, minimum width = \minwidth, minimum height = .5cm, fill = orange!20, text centered, anchor =north]

\tikzstyle{depthwise_block} = [draw, rectangle, minimum width=1.5cm, minimum height=1.5cm, fill=blue!20, text centered, shapes.geometric]

\tikzset{
    arr/.style={{Round Cap[]}-{Round Cap[]}, line width=1.5mm, shorten >=3.5mm, shorten <=3.5mm}
}

\usepgfplotslibrary{external}
\title{Signal-Based Malware Classification Using 1D CNNs}

\author*[1]{Jack Wilkie}
\author[2]{Hanan Hindy}
\author[1]{Ivan Andonovic}
\author[1]{Christos Tachtatzis}
\author[1]{Robert Atkinson}
\affil*[]{Jack Wilkie, jack.wilkie.2017@uni.strath.ac.uk}
\affil[1]{University of Strathclyde}
\affil[2]{Ain Shams University}

\begin{document}

\abstract{
  Malware classification is a contemporary and ongoing challenge in cyber-security: modern obfuscation techniques are able to evade traditional static analysis, while dynamic analysis is too resource intensive to be deployed at a large-scale. One prominent line of research addresses these limitations by converting malware binaries into 2D images by heuristically reshaping them into a 2D grid before resizing using Lanczos resampling. These images can then be classified based on their textural information using computer vision approaches. While this approach can detect obfuscated malware more effectively than static analysis; the process of converting files into 2D images results in significant information loss due to both quantisation noise, caused by rounding to integer pixel values, and the introduction of 2D dependencies which do not exist in the original data. This loss of signal limits the classification performance of the downstream model. This work addresses these weaknesses by instead resizing the files into 1D signals which avoids the need for heuristic reshaping, additionally these signals do not suffer from quantisation noise due to being stored in a floating-point format. It is shown that existing 2D CNN architectures can be readily adapted to classify these 1D signals for improved performance. Furthermore, a bespoke 1D convolutional neural network, based on the ResNet architecture and squeeze-and-excitation layers, was developed to classify these signals and evaluated on the MalNet dataset. It was found to achieve state-of-the-art performance on binary, type, and family level classification with F1 scores of \emph{0.874}, \emph{0.503}, and \emph{0.507}, respectively, paving the way for future models to operate on the proposed signal modality.
  }


\maketitle

\section{Introduction}
\label{sec:introduction}
Malware detection remains a critical challenge in cyber-security. While signature-based classification is effective for known malware, it relies on handcrafted features which are time-consuming and require expert knowledge to develop~\citep{247696, 6997496}. Furthermore, small changes to a malware's source code can greatly alter the compiled binary, allowing signature-based detection to be readily evaded~\citep{10.1007/978-3-540-70500-0_25}. While dynamic analysis provides a more in-depth inspection, its resource intensive nature limits its applicability at scale, posing significant challenges for cybersecurity defences that require rapid and accurate malware assessments.

Recent advances in machine learning have offered promising avenues for enhancing malware classification. Specifically, the transformation of malware binaries into visual image representations, known as byteplots, has emerged as a novel method that leverages computer vision techniques to classify malware samples based on the byteplot's textural information~\citep{10.1007/978-3-540-85933-8_1, 10.1145/2016904.2016908}. This approach boasts numerous advantages, removing the need to develop hand-crafted features and malware signatures, moreover the malware images are quick to generate allowing these methods to be deployed at scale. Significantly, this approach has enabled deep learning classification models which have been shown to be robust to both polymorphic malware variants and obfuscation techniques such as section reordering and file packing~\citep{10.1145/2016904.2016908, 10.1145/2046684.2046689,10040557}.

However, this approach is not without its limitations. The process of converting malware files into 2D images introduces significant information loss due to quantisation noise when rounding to integer pixel values and, more significantly, the artificial introduction of 2D spatial dependencies, which can distort the original binary structure and degrade the performance of downstream classification models.

Addressing these limitations, this work proposes an innovative approach that resizes the malware binaries into 1D signals instead of 2D images. This method not only maintains the integrity of the original data structure, but also retains more of the binary's information, improving the signal-to-noise ratio of the resized signal. It is shown that existing 2D models can be adapted to operate on 1D data for improved performance with no additional parameters or computation. Furthermore, a bespoke Convolutional Neural Network~(CNN) architecture is proposed to classify the 1D signal data. This architecture is adapted from the pre-activation ResNet architecture~\citep{he2016identity}, however, also incorporates squeeze-and-excitation~(SE) layers~\citep{squeeze_excite} and the GELU activation function~\citep{hendrycks2023gaussian}. 

While the proposed pipeline can be applied to arbitrary file binaries, this work focuses on the classification of Android DEX files due to the availability of a large-scale dataset for training and evaluation, and the substantial adoption of third-party applications in the Android ecosystem. Despite this, the proposed signal-based classification method is experimentally shown to be more effective than the equivalent image-based approaches on Windows' EXE binaries, making it an attractive replacement to image-based models.

Comprehensive evaluation on the MalNet dataset~\citep{10.1145/2016904.2016908} demonstrates that the proposed approach not only surpasses 2D image-based models but also achieves state-of-the-art performance on binary, type, and family level malware classification. This work paves the way for new malware classification systems which rely on 1D signal-based representations.

The core contributions of this work can be summarised as follows: \begin{enumerate}
    \item A resizing approach is proposed which keeps the file binaries in a 1D format, improving the signal-to-noise ratio when compared to resizing as a 2D image. Additionally, by storing the signals in a floating-point format quantisation error is avoided.
    
    \item A procedure to adapt arbitrary 2D CNNs to operate on the 1D signal representations is proposed. The 1D models were compared to their equivalent 2D models in a malware classification task on the MalNet and the Microsoft malware classification datasets. It was found that for an equal number of parameters and compute the 1D models outperform their 2D counterparts.
    
    \item A bespoke 1D CNN model is proposed to classify the malware signals. The model is based on the pre-activation ResNet architecture with GELU activation functions and SE layers. The proposed model outperforms state-of-the-art approaches in binary, type, and family classification on the MalNet dataset.

  \end{enumerate}

This work is organised as follows: related works are outlined in Section~\ref{sec:related_works}, including previous approaches to malware classification, byte plot images, and previous approaches utilising 1D convolutions for malware classification. Section~\ref{sec:proposed_approach} provides overview and explanation of the proposed methods including generating the signal files from binaries, and the novel CNN architecture used. The proposed approach is experimentally evaluated in Section~\ref{sec:experimental_results}, with additional results and ablations being provided in Section~\ref{sec:ablations}. Finally, relevant discussion and conclusions are provided in Section~\ref{sec:conclusions}.

\section{Related Works}
\label{sec:related_works}
In this section, a summary of previous works is given. Existing malware classification approaches are described in Section~\ref{sec:related_cls}, where methods such as static analysis, dynamic analysis, and machine learning approaches are detailed. Section~\ref{sec:related_byteplot} covers the history and application of byteplot representations, including their uses in manual binary inspection and as an input to machine learning models. Finally, Section~\ref{sec:related_1d_models} describes other works which have used 1D convolutions for malware classification.

\subsection{Malware Classification}
\label{sec:related_cls}

The classification of malware involves the extraction of distinctive features that characterise the contents and functionalities of malicious software. These features are typically gathered through static or dynamic analysis methods. Static analysis extracts features directly from the malware files, such as strings, n-grams, and byte-entropies~\citep{ABUSITTA2021102828}.  Conversely, dynamic analysis, while more time-intensive, provides deeper insights into the malware's functionalities. This method entails running the malware within a controlled virtual environment to observe behaviours such as network activity, permissions, and API calls. This approach is particularly effective against malware that employs obfuscation techniques to evade detection~\citep{ABUSITTA2021102828}. 

Signature-based detection remains a prevalent method in antivirus engines, where human analysts identify common properties among malware samples to define a malware signature. Such signatures may include filenames, byte sequence regular expressions, or printable strings~\citep{Schultz200138}. Despite the high precision of signature-based systems, they are susceptible to evasion through anti-analysis techniques including obfuscation, polymorphism, packing, and code reordering. These techniques alter the malware binaries substantially with minimal changes to their functionalities, thereby evading traditional detection methods~\citep{10.1109/COMSNETS53615.2022.9668396}.

In response to the limitations of signature-based detection, researchers have increasingly turned to machine learning classifiers capable of generalisation. Various gradient free methods have been applied such as Support Vector Machines~(SVMs)~\citep{Rezende2018MaliciousSC}, Markov Chains~\citep{10.1145/2381896.2381900}, and K-Nearest Neighbours Classifiers~(KNNs)~\citep{10.1145/2016904.2016908, Rezende2018MaliciousSC}, as well as deep models such as CNNs~\citep{malnet_dataset} and Vision Transformers~(ViTs)~\citep{Seneviratne_2022}. One prevalent method involves restructuring and resizing the binary data into a 2D image which can then be classified using computer vision approaches, this has numerous advantages such as being quick to generate and being resistant to obfuscation~\citep{10.1145/2016904.2016908}.

\subsection{Byteplot Visualisation}
\label{sec:related_byteplot}

The visualisation of binaries as byteplots was initially proposed as a tool for security analysts, wherein each byte was represented as a pixel with an intensity corresponding to the integer value of the byte~\citep{10.1007/978-3-540-85933-8_1}. These byteplots were displayed in a window of fixed width, allowing users to scroll through the entire binary file. It was qualitatively demonstrated that properties of the binary could be inferred from the textural information of the plots. For instance, network traffic of varying length packets manifested as an irregular texture, and hidden messages could be discerned in MP3 files. This enabled analysts to identify regions of interest without the need for manual examination of the binaries. Subsequent research extracted statistical information from byteplots and utilised it to automatically label primitive data types and sections within a binary file~\citep{CONTI2010S3}.

Byteplots have since found utility in computer vision-based classification approaches, where binary files are reshaped into a 2D grid. The width of this grid varies depending on the size of the binary, according to a heuristic resizing rule. The resulting 2D grid is then resized into a 2D image suitable for classification~\citep{10.1145/2016904.2016908}. Initially, filters were manually designed to extract feature vectors from these images, which could then be classified using SVMs~\citep{Rezende2018MaliciousSC} or KNNs~\citep{10.1145/2016904.2016908}. However, the success of CNNs on byteplot images has obviated the need for manual feature extraction~\citep{Rezende2018MaliciousSC, 8328749, Llaurad2016ConvolutionalNN}.

While it is common to generate greyscale byteplot images using standard procedures, several variations have been proposed. One approach seeks to enhance the images by incorporating colour channels to encode semantic information~\citep{Gennissen2017GamutS}, or by encoding multiple bytes into a single pixel~\citep{huang2018r2d2}. Another variation suggests replacing the commonly used heuristic reshaping rule with a method that shapes the data into a 2D grid with a width determined by the length of the longest section in the binary. However, this results in significant padding in rows corresponding to smaller sections~\citep{chong2022classification}.

Due to the success of computer vision-based classification, byteplot representations have been widely adopted in research~\citep{10061076}. Notably, the MalImg dataset contains 9,339 byteplot images from 25 different families~\citep{10.1145/2016904.2016908}. Another dataset, MalNet, comprises over 1 million byteplot images extracted from the DEX files of Android APKs~\citep{malnet_dataset}. Additionally, converting binaries into byteplot images is a commonly used preprocessing step~\citep{8328749, Llaurad2016ConvolutionalNN, 9877977} in works utilising the Microsoft Malware Classification dataset~\citep{microsoft_malware_dataset}.
 
\subsection{1D CNNs for Malware Classification}
\label{sec:related_1d_models}

1D CNNs have been utilised for malware classification, though not by operating on a 1D signal representations. A typical approach involves extracting tabular features through static or dynamic analysis. Common features can include n-grams~\citep{Liu2017, paul2022nlp}, header-based features~\citep{electronics10040485}, structural entropy~\citep{https://doi.org/10.1111/coin.12521, 10061401}, and byte/opcode frequencies~\citep{Liu2017, https://doi.org/10.1111/coin.12521}. These features are then used as input for 1D CNNs, either as standalone classifiers~\citep{9912986} or within an ensemble of models~\citep{electronics10040485}.

1D CNNs have shown prominence in processing sequences derived from API calls or opcode features. Initially, data is collected as strings via disassembly~\citep{Llaurad2016ConvolutionalNN, 10.1145/3029806.3029823, paul2022nlp, 10.32604/cmes.2022.018492, 8766515, 9411822} or dynamic analysis~\citep{10.5121/csit.2021.110106}. These strings are then parsed to identify a discriminative subset of opcodes or API calls, creating a short sequence for each sample. The sequences are then encoded using methods such as one-hot encoding~\citep{10.5121/csit.2021.110106, Schofield2021, 10.1145/3029806.3029823}, look up tables~\citep{10.1145/3029806.3029823}, GloVe~\citep{10.32604/cmes.2022.018492}, word-2-vec~\citep{paul2022nlp, Llaurad2016ConvolutionalNN} or frequency based features~\citep{Schofield2021} to obtain a sequence of embeddings which can then be classified using a shallow CNN~\citep{Schofield2021, 10.1145/3029806.3029823, 10.32604/cmes.2022.018492, 10.5121/csit.2021.110106}, or convolutional layers followed by a recurrent model~\citep{8659358, 8766515, Zyout2023, 9411822}. However, these data processing methods are time-consuming, and often these models underperform compared to byteplot or handcrafted feature-based approaches~\citep{10.1145/3029806.3029823}.

Despite the prevailing preference of 2D byteplot images in machine learning based approaches on raw binaries, some works have shown evidence of 1D convolutions providing performance benefits. One work has shown that the convolutional filters in 2D CNNs trained on byteplot images show a preference towards row-based features with column based features being less impactful on the binaries predicted label~\citep{chong2022classification}. Additionally, it was found that by ensembling a 2D CNN with features extracted by a single 1D convolutional layer operating on flattened byteplots provided performance benefits over using a 2D model alone~\citep{chong2022classification}. Other works have also performed 1D convolutions on flattened byteplot images, finding performance benefits in eliminating the vertical dimension in the convolutions~\citep{10061401}. Although these works have used relatively shallow networks, it has been attempted to train a Vision Transformer on entire 1D binaries, however, the authors found this impractical due to the file sizes involved and instead switch to a 2D byteplot based approach~\citep{9877977}. Other attempts have dealt with this file size constraint by operating only on the first or last \emph{N} bytes in a file~\citep{8368693} or by identifying discriminate regions within a binary and extracting sub-sequences for classification~\citep{9665792}. This paper presents the first work to train deep 1D CNNs on entire binaries resized into 1D signals, marking a novel approach in machine learning based malware classification.


\section{Proposed Approach}
\label{sec:proposed_approach}

\begin{figure*}[t]
  \centering
  \resizebox{\linewidth}{!}{
    \begin{tikzpicture}[
      xscale=2.8,yscale=2.4,
      myarrow/.style={->,very thick,draw=mydarkgreen},
      mynode/.style={
          thick,draw=mydarkgreen,fill=mylightgreen,
          rectangle,rounded corners=4,align=center,
          minimum width=3.5cm, 
          minimum height=1.5cm   
      },
      mycap/.style={shorten <=-0.3,line cap=round}
    ]
      
      \node[mynode] (RB) {
      \raggedbottom 
      \textbf{Raw Binary} \\[1.5ex] 
      {[}01000111...{]}
  };

      \node[mynode, right = 0.5 of RB] (SIGNAL) {
          \textbf{Signal} \\ \\
          \includegraphics[height = 1 cm]{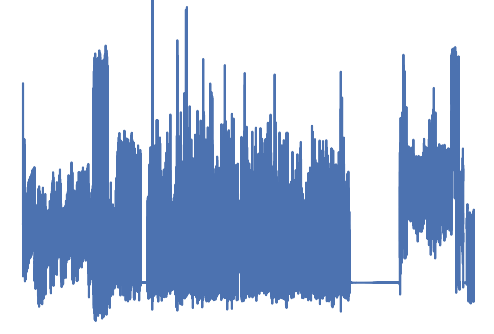}
      };
  
      \draw[myarrow] (RB) -- (SIGNAL);
      
      \node[mynode, right = 0.5 of SIGNAL] (1DCNN) {
          \textbf{1D CNN} \\ \\
          \begingroup
          \begin{tikzpicture}
            \node[draw, fill=white, minimum width=2cm, minimum height=1cm] (CNN) {\textbf{1D CNN}};
        
            \coordinate (arrow_start) at ([xshift=-0.5cm] CNN.west);
            \draw[->] (arrow_start) -- (CNN.west);
        
            \coordinate (arrow_end) at ([xshift=0.5cm] CNN.east);
            \draw[->] (CNN.east) -- (arrow_end);

        \end{tikzpicture}
        
          \endgroup
      };
  
      \draw[myarrow] (SIGNAL) -- (1DCNN);
  
      \node[mynode, above = 0.5 of 1DCNN] (2DCNN) {
          \textbf{2D CNN} \\ \\
          \scalebox{0.85}{\begin{tikzpicture}
            \fill[white] (0,0) rectangle (1.3,1);
            \draw[thick, fill=white] (-1, -0.5) -- (1, -0.5) -- (1, 0.5) -- (-1, 0.5) -- cycle; 
            \draw[thick, fill=white] (1, -0.5) -- (1.5, 0) -- (1.5, 1) -- (1, 0.5) -- cycle; 
            \draw[thick, fill=white] (-1, 0.5) -- (-0.5, 1) -- (1.5, 1) -- (1, 0.5) -- cycle; 
        
            \node[anchor = center] at (0, 0) {\textbf{2D CNN}};
        
            \draw[->] (-1.5, .25) -- (-1, .25);
        
            \draw[->] (1.25, .25) -- (1.85, .25);
        
        \end{tikzpicture}
        }
      };
  
      \draw[myarrow] (2DCNN) -- (1DCNN);
  
      \node[mynode, right = 0.5 of 1DCNN] (CLASS) {
          \textbf{File Classification} \\ \\
          \centering
          \includegraphics[width= 1 cm]{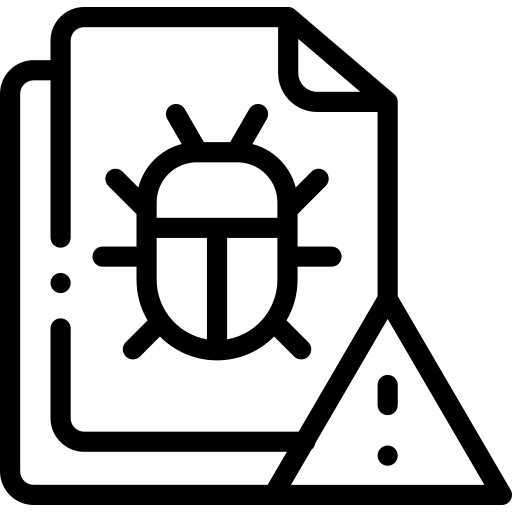}
      };
  
      \draw[myarrow] (1DCNN) -- (CLASS);
  
  \end{tikzpicture}
  
  } 
  \caption{High-level overview of the proposed system. File binaries are initially converted into 1D signals, which preserve more of the file information, before being classified using a 1D CNN. Additionally, existing 2D CNNs can be adapted to operate on the 1D signals by modifying the convolution kernels.}
  \label{fig:system_overview}
  \end{figure*}
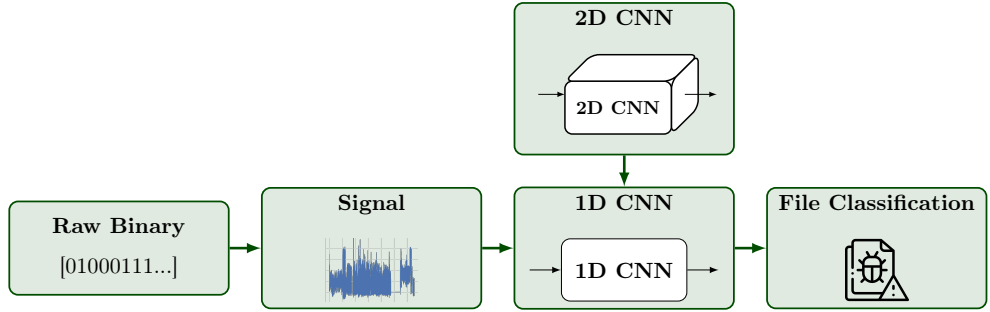

A high-level schematic of the proposed approach is provided in Figure~\ref{fig:system_overview}. This represents an approach to perform malware classification based on 1D signal representations of binary files. These representations are produced by resizing binaries into 1D signals using Lanczos resampling as described in Section~\ref{sec:data_resizing}. To classify the signal representations a methodology is presented in Section~\ref{sec:method_conv_conversion} to convert existing 2D CNNs to operate on 1D data by flattening the convolution kernels and squaring the stride values. This technique is applied to the ResNet model along with several modifications to develop a bespoke architecture adept at classifying the generated signals in Section~\ref{sec:proposed_arch}.

\subsection{Data Processing and Resizing}
\label{sec:data_resizing}

Instead of processing malware binaries into byteplot visualisations, this work converts them into signal representations. For the purposes of this work the term ``signal'' is used to refer to 1D inputs to machine learning models, in order to differentiate from 2D images. Initially, the binary file is transformed into an integer representation where each byte is represented by an integer in the range $[0,255]$, corresponding to the byte's numerical value when interpreted as an integer.

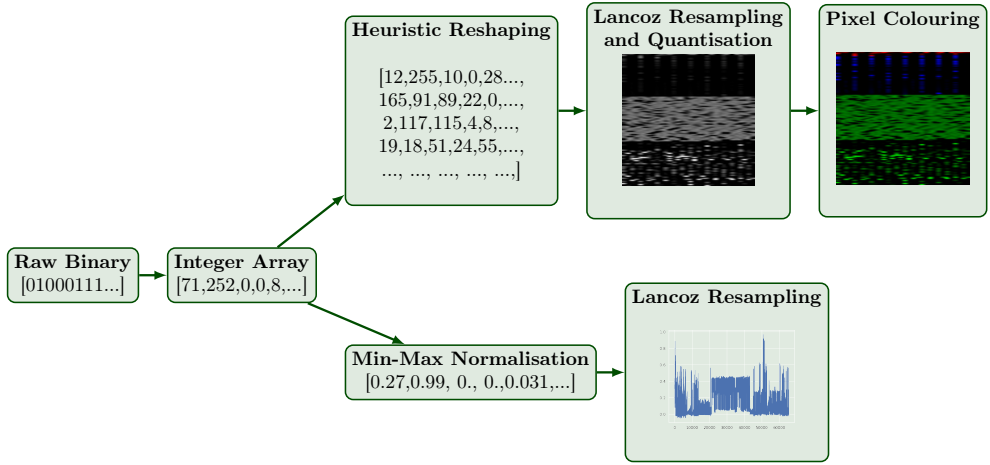
\begin{figure*}[t]
  \centering
  \resizebox{\linewidth}{!}{
  \begin{tikzpicture}[
      xscale=2.8,yscale=2.4,
      myarrow/.style={->,very thick,draw=mydarkgreen},
      mynode/.style={thick,draw=mydarkgreen,fill=mylightgreen,rectangle,rounded corners=4,align=center},
      mycap/.style={shorten <=-0.3,line cap=round}
    ]

      \node[mynode] (BR) {
      \textbf{Raw Binary} \\
      {[}01000111...{]}};
      
      \node[mynode, right = 0.5 of BR] (IR) {
      \textbf{Integer Array} \\ 
      {[}71,252,0,0,8,...{]}};
      \draw[myarrow] (BR) -- (IR);
  
      \node[mynode, above right = 0.75 and 0.5 of IR] (IMSHAPE) {
      \textbf{Heuristic Reshaping} \\ \\
      {[}12,255,10,0,28..., \\
      165,91,89,22,0,..., \\
      2,117,115,4,8,..., \\
      19,18,51,24,55,..., \\
      ..., ..., ..., ..., ...,{]} \\
      };
      \draw[myarrow] (IR) -- (IMSHAPE);
  
      \node[mynode, right = 0.5 of IMSHAPE] (IMSIZE) {
      \textbf{Lancoz Resampling} \\ 
      \textbf{and Quantisation} \\
      \includegraphics[width= 2.5 cm]{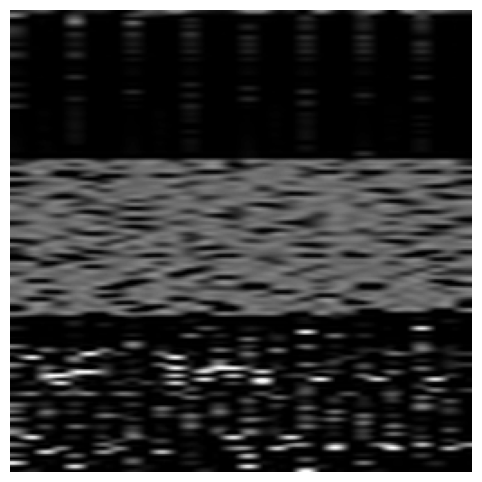} \\
      };
      \draw[myarrow] (IMSHAPE) -- (IMSIZE);
  
      \node[mynode, right = 0.5 of IMSIZE] (IMCOL) {
      \textbf{Pixel Colouring} \\ \\
      \includegraphics[width= 2.5 cm]{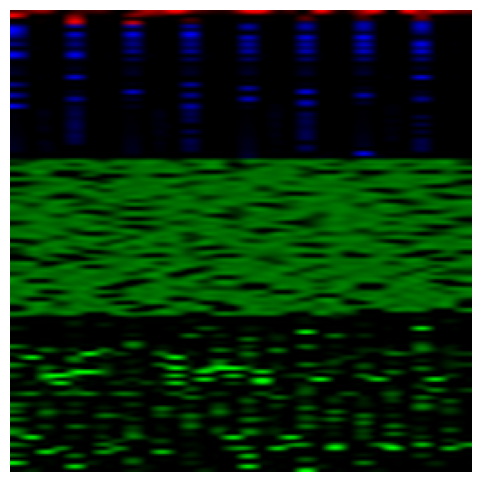} \\
      };
      \draw[myarrow] (IMSIZE) -- (IMCOL);
      
       \node[mynode, below right = 0.75 and 0.5 of IR] (SNORM) {
          \textbf{Min-Max Normalisation} \\
          {[}0.27,0.99, 0., 0.,0.031,...{]}};
      
      \draw[myarrow] (IR) -- (SNORM);
  
      \node[mynode, right = 0.5 of SNORM] (SSIZE) {
          \textbf{Lancoz Resampling} \\ \\
      \includegraphics[width= 2.5 cm]{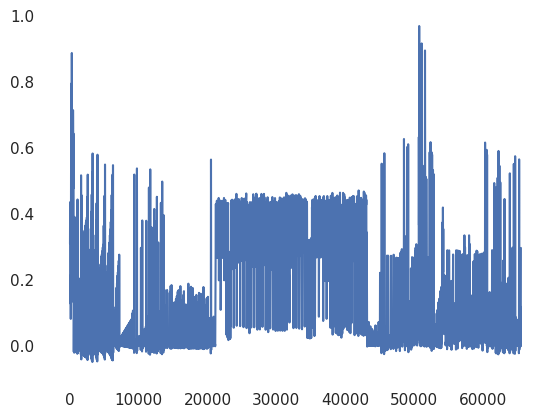} \\
      };
      
      \draw[myarrow] (SNORM) -- (SSIZE);
    
  \end{tikzpicture}
  }
  \caption{The signal processing procedure use to generate both byteplot images (top path) and signal representations (bottom path). The signals are generated in a similar manner to bytplot images but do not require the reshaping or quantisation steps which introduce unnecessary noise when generating byteplot images.}
  \label{fig:sig_processing}
  \end{figure*}

As depicted in Figure~\ref{fig:sig_processing}, generating a byteplot image involves arranging integer values into a 2D grid using a heuristic resizing rule. However, the proposed data processing method diverges at this point by employing min-max normalisation. Here, each integer value is divided by 255, resulting in a floating-point representation in the range $[0,1]$. By maintaining the data's original 1D structure the signal representations avoid the introduction of 2D spatial dependencies which result in lower quality resizing and diminished downstream classification performance.

Similar to byteplot visualisations, the Lanczos filter is used to resample the data. This is due to its ability to preserve textural information in malware binaries useful for classification. Although signals can be resampled to arbitrary size, this work resamples to a length of $65536$, equal to the number of pixels in a $256\times256$ image. This allows for fair comparison between the proposed approach and existing image-based classifiers. Post-resizing, the signal remains in floating-point format, which avoids the quantisation loss suffered by byteplot representations.

\begin{figure}[t]
  \centering
  \resizebox{\linewidth}{!}{
  \begin{tikzpicture}[
      xscale=2.8,yscale=2.4,
      myarrow/.style={->,very thick,draw=mydarkgreen},
      mynode/.style={thick,draw=mydarkgreen,fill=mylightgreen,rectangle,rounded corners=4,align=center},
      mycap/.style={shorten <=-0.3,line cap=round}
    ]
    
      \node[mynode] (ex1) {
      \includegraphics[width= 1.6 in ]{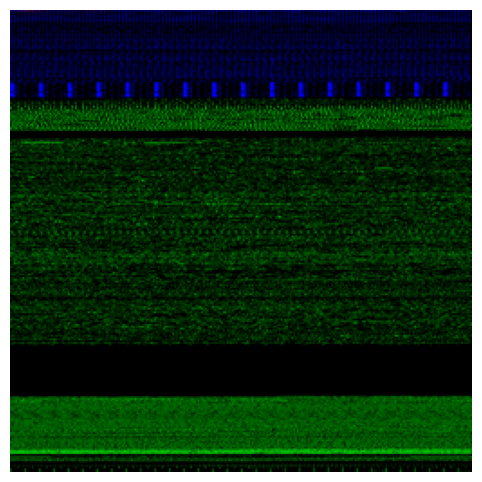} 
      \includegraphics[width= 1.6in]{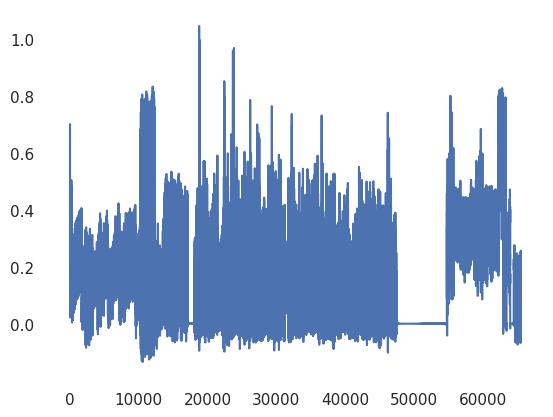}
      \\
      Benign \\
      Benign
      };

      \node[mynode, right = 0.5 of ex1] (ex2) {
      \includegraphics[width= 1.6 in ]{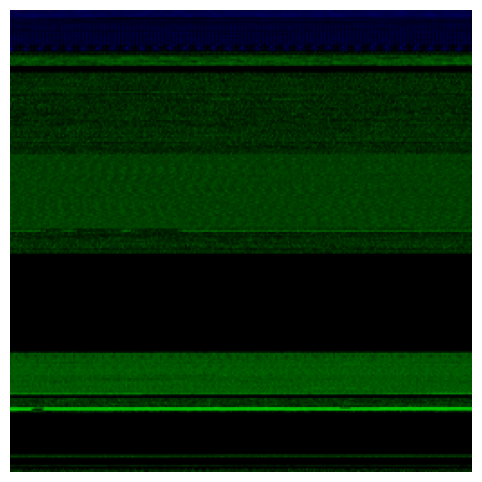} 
      \includegraphics[width= 1.6in]{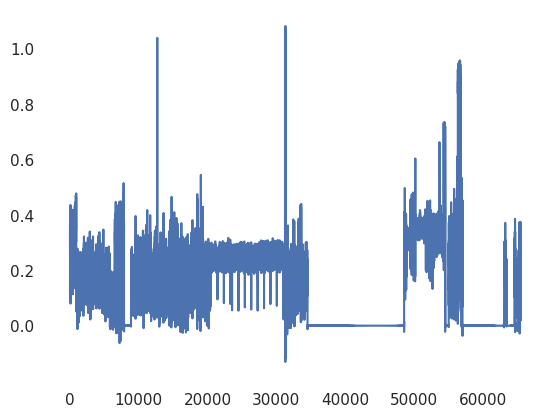}
      \\
      Adware \\
      Kyview
      };
  
      \node[mynode, below = 0.5 of ex1] (ex3) {
      \includegraphics[width= 1.6 in ]{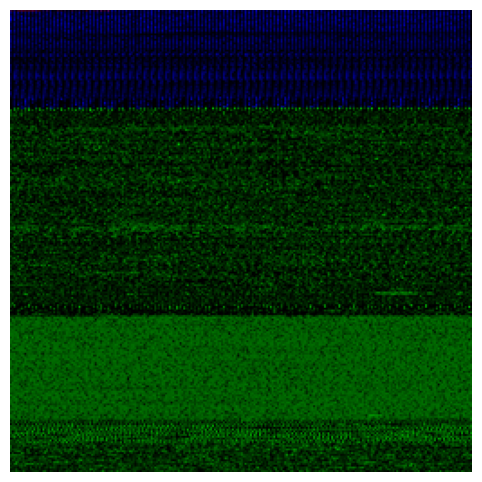} 
      \includegraphics[width= 1.6in]{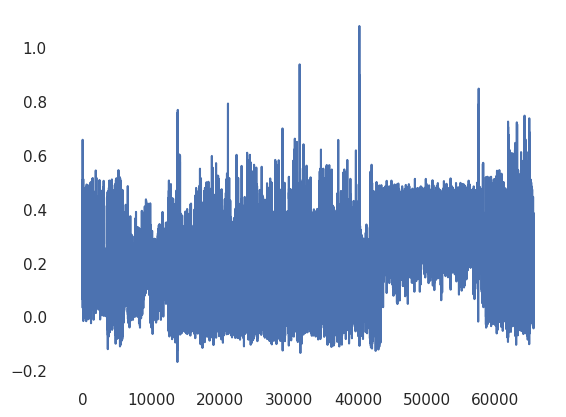}
      \\
      Trojan \\
      GingerMaster
      };
  
      \node[mynode, right = 0.5 of ex3] (ex4) {
      \includegraphics[width= 1.6 in ]{2d_riskware_artemis_0011672318B8644607BE27D3735EBF82A1EDBB393BD54B2359D0D85A7D0715A7.png} 
      \includegraphics[width= 1.6in]{1d_riskware_artemis_0011672318B8644607BE27D3735EBF82A1EDBB393BD54B2359D0D85A7D0715A7.png}
      \\
      Riskware \\
      Artemis
      };
  \end{tikzpicture}
  }
  \caption{Comparison of byteplot images with the corresponding signal representation. Visual similarity can be seen between the representations, most notably on areas of padding which appears as  black regions in the image and $0$ regions in the signals. The captions show the type of malware (top) and its family (bottom).}
  \label{fig:sig_comparison}
  \end{figure}

In contrast to byteplot visualisations, this work does not encode binary section information as multiple channels; instead, a single channel is used. While encoding information as colour channels is beneficial for visualisation purposes, it does not enhance the performance of the downstream classifier despite resulting in additional model complexity and data storage requirement~\citep{malnet_dataset}. The byteplot visualisation of samples from the MalNet dataset and their corresponding signal representations are shown in Figure~\ref{fig:sig_comparison}. Correlation can be seen between the byteplot image's pixel intensity and the corresponding signal values. Notably regions of padding appear as black regions in byteplot images and long $0$ regions in the signal representation. Before model training, the mean and standard deviation of the signal are calculated on the training set, enabling z-normalisation of the signal.

\subsection{Converting Convolutions to One Dimension}
\label{sec:method_conv_conversion}
Typical 2D CNN classifiers are unable to process the proposed 1D signal representations of file binaries, necessitating an architecture able to operate on these signals. Existing approaches in other domains have attempted to adapt CNNs to 1D data by either preserving the convolution size along a single dimension---for instance, converting a 2D convolutional layer with a $3 \times 3$ kernel into a 1D convolutional layer with a kernel size of $3$~\citep{10.1088/2632-2153/aced7f}---or by employing custom architectures with kernel sizes tailored to the specific characteristics of the input data~\citep{10.3390/s22083094}. However, these adaptations do not result in 1D architectures equivalent to their 2D counterparts, with discrepancies in model structure, parameter count, and computational requirements when compared to their 2D equivalents.

\begin{figure}[t]
  \centering
  \resizebox{\linewidth}{!}{
    \includegraphics{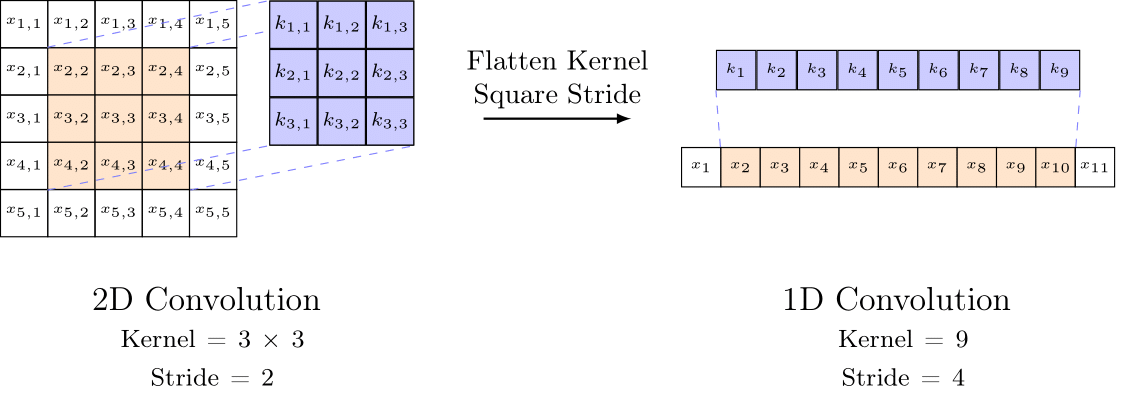}
  }
  \caption{Conversion of a 2D CNN Kernel into a 1D CNN kernel. By flattening the kernels and squaring the stride values across a 2D CNN, an equivalent 1D CNN can be found with the same number of parameters and compute requirement. Here the kernel is represented by $k$, and the input is given as $x$.}
  \label{fig:cnn_conversion}
\end{figure}

This work instead proposes a novel approach to convert 2D CNNs into 1D CNNs that preserves the network architecture and key properties such as parameter count and computational cost. As illustrated in Figure~\ref{fig:cnn_conversion}, instead of modifying kernel sizes arbitrarily, the kernels are flattened, maintaining the same number of parameters, except across a single dimension. Additionally, the stride values are squared preserving the downsampling ratio in convolutional layers (a stride of $2$ across $2$ dimensions results in a total downsampling ratio of $4$). By performing this transformation on all 2D convolutions in a model, existing CNN architectures can readily be adapted to the proposed signal representations. The 1D models show improved performance for no extra parameter or compute cost over their 2D counterparts.

\subsection{Model Architecture}
\label{sec:proposed_arch}

\begin{figure}[t]
  \centering
  \resizebox{\linewidth}{!}{
      \begin{tikzpicture}
    
          \node[rectangle, align=center] (input) {
              \tikz \draw[thick] plot[domain=-pi:pi, samples=100] (\x * -0.2,{sin(\x r)*0.5});
              \\
              Input Signal    
              \\
              (Bx1xL)
          };
          
          \node[conv, right = .5 of input.east, anchor = north, rotate = 90](conv1) {Conv1D};
            
          \draw[->] (input) to (conv1);

          \node[norm, right = 0. of conv1.south, anchor = north, rotate = 90](norm1) {GroupNorm};
          \node[activation, right = 0. of norm1.south, anchor = north, rotate = 90](act1) {GELU};
          
          \node[conv, right = 0.0 of act1.south, anchor = north, rotate = 90](conv2) {Conv1D};
          \node[norm, right = 0. of conv2.south, anchor = north, rotate = 90](norm2) {GroupNorm};
          \node[activation, right = 0. of norm2.south, anchor = north, rotate = 90](act2) {GELU};
          
          \node[conv, right = 0.0 of act2.south, anchor = north, rotate = 90](conv3) {Conv1D};

          \node[pooling, right = . of conv3.south, anchor = north, rotate = 90](pool) {Max Pooling};

          \node[below = 0.1 of conv1.north west, anchor = north](bracket1_start){};
          \node[below = 0.1 of pool.south west, anchor = north](bracket1_end){};
          \draw[decorate,decoration={brace,amplitude=10pt, mirror}] (bracket1_start) -- (bracket1_end) node[midway,below=10pt] {\emph{ResNet Stem}};

          \node[container, right = .5 of pool.south, anchor = north, rotate = 90](stage1) {Stage Block 1};
          \draw[->] (pool.south) to (stage1);
          
          \node[container, right = .5 of stage1.south, anchor = north, rotate = 90](stage2) {Stage Block 2};
          \draw[->] (stage1.south) to (stage2);
          
          \node[container, right = .5 of stage2.south, anchor = north, rotate = 90](stage3) {Stage Block 3};
          \draw[->] (stage2.south) to (stage3);
          
          \node[container, right = .5 of stage3.south, anchor = north, rotate = 90](stage4) {Stage Block 4};
          \draw[->] (stage3.south) to (stage4);

          \node[below = 0.1 of stage1.north west, anchor = north](bracket2_start){};
          \node[below = 0.1 of stage4.south west, anchor = north](bracket2_end){};
          \draw[decorate,decoration={brace,amplitude=10pt, mirror}] (bracket2_start) -- (bracket2_end) node[midway,below=10pt] {\emph{ResNet Stage}};


          \node[norm, right = .5 of stage4.south, anchor = north, rotate = 90](norm3) {GroupNorm};
          \draw[->] (stage4.south) to (norm3);
          
          \node[activation, right = 0. of norm3.south, anchor = north, rotate = 90](act3) {GELU};

          \node[pooling, right = 0. of act3.south, anchor = north, rotate = 90](pool1) {Average Pooling};
          
          \node[conv, right = 0. of pool1.south, anchor = north, rotate = 90, fill=orange!20](linear1) {Linear Layer};

          \node[below = 0.1 of norm3.north west, anchor = north](bracket3_start){};
          \node[below = 0.1 of linear1.south west, anchor = north](bracket3_end){};
          \draw[decorate,decoration={brace,amplitude=10pt, mirror}] (bracket3_start) -- (bracket3_end) node[midway,below=10pt] {\emph{Classification Head}};

          \node(output)[right = 0.5 of linear1.south]{};
          \draw[->] (linear1) to (output);

      \end{tikzpicture}
  }
\caption{Architecture of the proposed 1D ResNet. The model is based on the ResNet architecture with a deep stem, however, 2D convolutions are replaced with 1D Convolutions. Additionally, the kernel is flattened and the stride is squared for each convolution.}
\label{fig:resnet_architecture}
\end{figure}
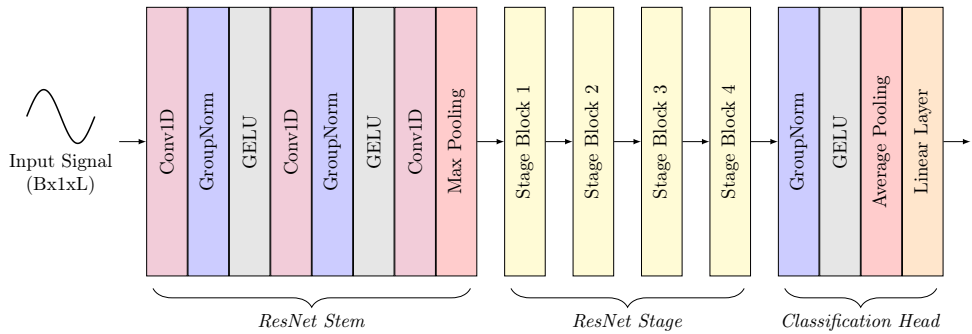

To demonstrate the efficacy of malware classification using signal representations, the bespoke 1D CNN architecture shown in Figure~\ref{fig:resnet_architecture} was developed by converting the 2D ResNet architecture~\citep{he2015deep} to operate on 1D signals. To modernise the architecture, the ReLU activation functions were replaced with GELU and squeeze-and-excitation~(SE) layers were introduced in the PRE-SE configuration~\citep{squeeze_excite}. The network has three stages: the stem, which performs initial feature extraction whilst spatially downsampling the input signal; the stage, which applies residual learning to progressively learn a feature map; and the classification head, which predicts the binary's class label based on the extracted features.

The model's stem utilises the deep architecture introduced by the ResNet-D models~\citep{he2018bag}, employing a hierarchical sequence of convolutional, normalisation, and downsampling operations to extract features from the input signal whilst spatially downsampling it to improve downstream efficiency. It consists of three convolutions: two initial convolutions, each followed by Group Normalisation~(GroupNorm) and a GELU activation function; and a third convolution which is followed by max pooling. The first convolution employs a stride of 4, reducing the input signal’s length by a factor of four whilst projecting it to 32 channels. The second convolution maintains the spatial dimensions and channel count, refining feature representations through additional non-linear transformations. A final convolutional layer then expands the channel depth to 64, followed by a max pooling operation with a stride of 4, further downsampling the signal length by a factor of four. A kernel size of nine is used across the max pooling and convolutions in the stem.

\begin{figure}[t]
  \centering
  \resizebox{\linewidth}{!}{
  \begin{tabular}{cc} \footnotesize
  
\begin{tikzpicture}

  \node(input){};
  
  \node[norm, below = 0.5 of input.south] (norm0){GroupNorm};
  \node[activation, below = 0.0 of norm0.south] (act0){GELU};
  \node[container, below = 0.0 of act0.south] (se1){Squeeze Excitation};            
  \node(res_start)[below = 0.5 of se1.south]{};
  \draw[->] (input) to (norm0);

  \node[conv, below = 0.5 of res_start.south] (conv1){Pointwise Conv1D};
  \draw[->] (se1) to (conv1);
  
  \node[norm, below = 0.0 of conv1.south] (norm1){GroupNorm};
  \node[activation, below = 0.0 of norm1.south] (act1){GELU};

  \node(mid_point)[below = .5 of act1.south]{};
  
  \node[conv, below = 1.25 of mid_point.south] (conv2){Depthwise Conv1D};
  \draw[->] (act1) to (conv2);

  \node[norm, below = 0.0 of conv2.south] (norm2){GroupNorm};
  \node[activation, below = 0.0 of norm2.south] (act2){GELU};

  \node[conv, below = 0.5 of act2.south] (conv4){ Pointwise Conv 1D};
  \node[mul, below = 0.0 of conv4.south] (add){Pointwise Addition};
  \draw[->] (act2) to (conv4);

  \node(output)[below = 0.5 of add.south]{};
  \draw[->] (add) to (output);

  \node[pooling, right = .5 of mid_point.east] (pool1){Average Pooling};
  \node[conv, below = 0.0 of pool1.south] (conv3){ Pointwise Conv 1D};
  
  \draw[-] (res_start.north) -| (pool1.north);
  \draw[->] (conv3.south) |- (add.east);
  
\end{tikzpicture}

  \hspace{1cm} &
  
\begin{tikzpicture}

  \node(input){};
  
  \node(res_start)[below = 0.5 of input.south]{};

  \node[pooling, below = 0.5 of res_start.south] (gap){Channel Average Pooling};
  \draw[->] (input) to (gap);

  \node(res_mid)[left = 0.5 of gap.west]{};

  \node[downsample, below = 0.5 of gap.south] (down1){Linear Downsampling};
  \draw[->] (gap) to (down1);
  
  \node[activation, below = 0.5 of down1.south] (act1){RELU};
  \draw[->] (down1) to (act1);
  
  \node[upsample, below = 0.5 of act1.south] (up1){Linear Upsampling};
  \draw[->] (act1) to (up1);
  
  \node[activation, below = 0.5 of up1.south] (act2){Sigmoid};
  \draw[->] (up1) to (act2);
  
  \node[mul, below = 0.5 of act2.south] (mul){Channelwise Multiplication};
  \draw[->] (act2) to (mul);

  \node(output)[below = 0.5 of mul.south]{};
  \draw[->] (mul) to (output);

  \draw[-] (res_start.north) -| (res_mid.north);
  \draw[->] (res_mid.north) |- (mul.west);
  
\end{tikzpicture}

  \end{tabular}
  }
  \caption{
      Architecture of the residual block. \textbf{Left:} The blocks use the pre-activation ResNet architecture with the addition of the GELU activation function and a squeeze-and-excitation layer. \textbf{Right:} architecture of a squeeze-and-excitation block used to provide input dependant convolution filters.
  }
  \label{fig:resblock_architecture}
  \end{figure}
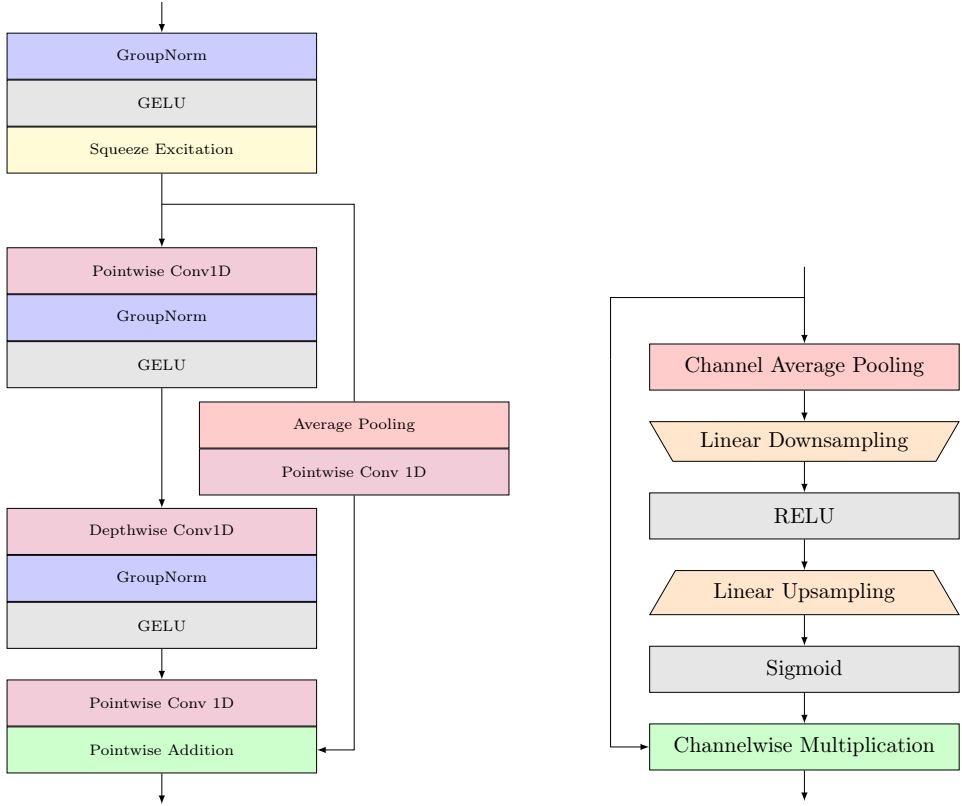

The ResNet stage consists of 4 stage blocks containing a series of the residual blocks shown in Figure~\ref{fig:resblock_architecture}~(left). These blocks are 1D adaptations of the pre-activation ResNetV2 architecture~\citep{he2016identity} with the aforementioned modernisations of the GELU activation and SE layers. Each residual block begins with GroupNorm, followed by a GELU activation and SE layer, after which the signal is processed through two parallel branches. The main branch employs separable convolutions, beginning with a pointwise convolution (followed by GroupNorm and GELU) which reduces the dimensionality of the feature map; before a depthwise convolution (also followed by GroupNorm and GELU) processes the signal in the lower dimensionality feature space. A second pointwise convolution then restores original number of channels. Meanwhile, the residual branch resamples the signal to match the shape of the main branch using average pooling (to resize the spatial dimension) and a pointwise convolution (to match the number of channels). The two branches are then reconciled using a pointwise addition. Importantly, the first residual block in each stage block downsamples the length of the signal by a factor of 4, using the stride parameter of the depthwise convolution, and doubles the number of channels using the pointwise convolutions. This progressive downsampling allows the model to learn semantic, hierarchical features as the signal propagates through the network.

The SE layers are used to provide channel-wise attention to the residual block. As illustrated in Figure~\ref{fig:resblock_architecture}~(right), global average pooling is used to get a compressed representation of the feature map in the current layer of the network. This pooled representation is then passed through an MLP, utilising the bottleneck architecture, to compute a scaling vector used gate the feature channels. By scaling the input to the convolutions, the model is able to exploit input dependant filters.

Finally, the classification head is used to predict the class of the input binary using the feature map learned by the ResNet stage. The extracted features are first normalised using GroupNorm. Following this, a GELU activation is used to allow the classification head to make predictions based on a non-linear combination of features. Finally, average pooling is used to eliminate the length dimension of the signal, with the remaining features being classified using a linear layer.

\section{Evaluation}
\label{sec:experimental_results}
This section experimentally evaluates this works primary contributions. Initially, the evaluation datasets and experimental procedures are detailed in Section~\ref{sec:datasets} and Section~\ref{sec:proceedure}, respectively. Next the performance impact of transitioning from a byteplot representation of a binary to a signal-based one is measured in terms of both signal loss and performance impact on downstream models in Section~\ref{sec:compression_loss}. Section~\ref{sec:conv_adaptation_ablation} compares various approaches to convert 2D convolutional filters to operate on 1D data. Finally, Section~\ref{sec:benchmarking} compares the proposed model to current SOTA approaches on the MalNet dataset in binary, type, and family level classification.

\subsection{Datasets}
\label{sec:datasets}

This work primarily uses the MalNet Image dataset~\citep{malnet_dataset}. This is the largest open source malware datasets containing 1,262,024 Android malware samples across 696 families and 47 types of malware, including benign samples. While the MalNet dataset is provided as byteplot images, the file hashes can be used to download the original APK files from Androzoo~\citep{Allix:2016:ACM:2901739.2903508}. After downloading the corresponding APKs the binary sequences from the DEX files were extracted and converted to a signal representation using the methodology described in Section~\ref{sec:data_resizing}.

To demonstrate the applicability of the proposed methods to other file formats the Microsoft malware classification dataset~\citep{microsoft_malware_dataset} was used. This dataset contains the raw binaries and dissassembled opcode of EXE files across nine malware families. While this dataset is used to show the applicability of the proposed approach to windows EXE files it is a much smaller scale than the MalNet dataset, containing only 10,868 samples, and thus has less sample variety. Consequently, the dataset has been largely solved with existing models achieving near perfect performance~\citep{GIBERT2020101873}. For this reason, the proposed 1D CNNs are compared to their equivalent 2D models on this dataset but are only compared to SOTA approaches on the Malnet dataset.

While other malware classification datasets exist, they cannot be used in this work as they do not provide the original file binaries. Although commonly used the MalImg~\citep{10.1145/2016904.2016908} and the VirusMNIST~\citep{noever2021virusmnist} datasets provide only the image representation of the files, making it impossible to convert the samples into a signal representation. Other datasets such as the EMBER dataset~\citep{2018arXiv180404637A} provide the data as a set of tabular features. All of these datasets are also significantly smaller than the MalNet dataset.

\subsection{Experimental Details}
\label{sec:proceedure}

Models trained on the MalNet dataset utilise the predefined 70/10/20 train, test, and validation sets for each task granularity. During training, models were evaluated at each epoch on the validation set, with the checkpoint exhibiting the best validation performance being selected for final evaluation on the test set. For the Microsoft Malware Classification dataset, due to the unavailability of public test set labels, the training set was split into an 80/20 train-validation split, with validation results being reported.

Two distinct training recipes were employed in this work. The standard recipe was derived from previous studies~\citep{malnet_dataset} and is used for a direct comparison with existing literature when evaluating the efficacy of transforming the data into a 1D signal-based representation. It trains the model for 100 epochs using the Adam optimiser with a learning rate of 0.001 and a batch size of 64. Additionally, an improved training recipe was introduced to compare the proposed model's performance against current state-of-the-art~(SOTA) approaches. The improved recipe incorporates several enhancements, including linearly scaling class balancing strength by a factor of 0.5, applying weight decay with strength 0.005, as well as label smoothing with an alpha value of 0.1, and using a warm-up cosine schedule for the learning rate. In the improved training recipe, models are initially trained using the stated hyperparameters for 50 epochs. Subsequently, the class balancing weights are removed, and the model is fine-tuned for a further 10 epochs with a learning rate which is reduced by a factor of 10. This improved recipe allows models to achieve improved performance despite training for fewer epochs.

In the remainder of this work ResNet models trained on signal data are referred to as ResNet1D models to distinguish them from traditional image-based ResNets. The standard notation is retained to denote the ResNet's size; for example, ResNet1D152 refers to a ResNet152 model adapted for signal data. The original ResNet block is denoted as the V1 architecture~\citep{he2015deep}, while the bottleneck architecture is referred to as V1.5~\citep{he2015deep}, and the pre-activation architecture is labelled as V2~\citep{he2016identity}. SE denotes the inclusion of squeeze-and-excitation modules, and D indicates the use of a deep ResNet stem. For instance, a ResNet1DV2-152D-SE model refers to a ResNet152 model utilizing the pre-activation architecture, trained on signal data, and incorporating both a deep stem and squeeze-and-excitation modules.

\subsection{Loss of Information in 2D Images}
\label{sec:compression_loss}
In this section, the effect of resizing binaries into 1D signals as opposed to 2D images is explored. To provide a quantitative comparison of the resizing methods, the mean squared error~(MSE) and signal-to-noise ratio~(SNR) was measured at each stage of the resizing process.

To assess resizing noise, the binaries were resized to a length of $65536$ for signals or reshaped into a grid and resized to $256\times256$ for images. The downsampled binary was then upsampled back to its original size, with the MSE and SNR being calculated between the original and the resized binary. Quantisation noise was measured by first resizing the binary in either 1D or 2D and then quantising to the nearest integer, comparing quantised signal to the resized signal. The quantised signals were subsequently upsampled back to their original sizes and compared to the original binaries to measure the total noise incurred from the combination of resizing and quantisation.

The mean resizing and quantisation noise for both images and signals is presented in Table~\ref{tab:malnet_noise_sources} for the MalNet dataset and in Table~\ref{tab:microsoft_noise_sources} for the Microsoft Malware Classification dataset. The results indicate that resizing noise is more significant when the data is resized into 2D images. Although on the Microsoft Malware dataset quantising signals introduces more noise than quantising image representations, it should be noted that the signal representations are not quantised when used to train a model, thus only resizing noise is suffered. Consequently, the total noise in the image representations is the cumulation of resizing and quantisation noise, whereas for signals, it is only resizing noise. The results demonstrate that for both Android APKs and Windows' EXE files, the proposed signal representations introduce significantly less noise during preprocessing.

\begin{table}[t]

  \caption{Sources of noise in the binary resizing process on the MalNet dataset. }
  \label{tab:malnet_noise_sources}
  \centering
  \begin{tabular}{lcc}
  
  \hline 
  Signal Loss Type & SNR (dB) & MSE \\
  \hline
  
  Image Resizing & -17.6461 & 3757.7163 \\
  Image Quantisation & 35.9104 & 1.2992 \\
  Image Resizing + Quantisation & -17.6439 & 3755.8519\\
  Signal Resizing & -16.9099 & 3238.4049 \\
  Signal Quantisation  & 40.2085 & 1.1551 \\
  Signal Resizing + Quantisation  & -16.8438 & 3240.5584 \\
  \hline

  \end{tabular}
  \end{table}

\begin{table}[t]

  \caption{Sources of noise in the binary resizing process on the Microsoft Malware Dataset. }
  \label{tab:microsoft_noise_sources}
  \centering
  \begin{tabular}{lcc}
  
  \hline 
  Signal Loss Type & SNR (dB) & MSE \\
  \hline
  
  Image Resizing & -19.3484 & 4937.4840 \\
  Image Quantisation & 44.6832 & 0.7755 \\
  Image Resizing + Quantisation  & -19.3483 & 4937.3816 \\
  Signal Resizing & -18.5801 & 4133.1473 \\
  Signal Quantisation  & 42.2795 & 3.2579 \\
  Signal Resizing + Quantisation  & -18.5871 & 4138.1813 \\
  \hline

  \end{tabular}
  \end{table}

Reduced preprocessing noise does not necessarily guarantee improved performance when training a classification model. To evaluate this, the performance of various 2D CNNs, trained on images, were compared to their 1D equivalent models, trained on signal representations. The performance of both sets of models on the MalNet type dataset is shown in Table~\ref{tab:malnet_resnet_comparison} with the comparison being extended to the Microsoft Malware Classification dataset in Table~\ref{tab:microsoft_resnet_comparison}. It is evident that despite having the same computational requirements and number of parameters, the 1D models outperform their 2D equivalent models, with each model achieving a higher F1 score when trained on 1D signals across both Android APK and Windows' EXE data.

\begin{table*}[t]
  \centering
  
  \caption{Comparison between 1D and 2D CNNs on type level classification when trained with identical hyperparameters.}
  
  \label{tab:malnet_resnet_comparison}
  \centering 
  \resizebox{\linewidth}{!}{\begin{tabular}{lccccc}
  
  \hline 
  Model & F1 Score & Precision & Recall & Parameters (M) & Compute (GFLOPs) \\
   
  \hline 
  ResNet18 & .467 & .556 & .424 & 11.2 & 2.3  \\
  ResNet50 & .479 & .566 & .441 & 23.6 & 9.9 \\
  DenseNet121 & .471 & .558 & .428 & 7.0 & 3.6 \\
  Densenet169 & .477 & .573 & .433 & 12.6 &  4.3\\
  MobileNetV2(x.5) & .460 & .547 & .424 &  0.7 & 0.1   \\
  MobileNetV2(x1) & .453 & .527 & .419 & 2.28 & 0.4 \\
  
  \midrule
  ResNet1D18 & .482 & .598 & .432 & 11.2 & 2.3  \\
  ResNet1D50 & .486 & .585 & .443 & 23.6 & 9.9 \\
  DenseNet1D121 & .499 & .574 & .466 & 7.0 & 3.6 \\
  Densenet1D169 & .488 & .562 & .452 & 12.6 &  4.3\\
  MobileNet1DV2(x.5) & .482 & .548 & .458 &  0.7 & 0.1   \\
  MobileNet1DV2(x1) & .481 & .552 & .444 & 2.28 & 0.4 \\
  
  \hline
  \end{tabular}}
  \end{table*}

  \begin{table}[t]

    \caption{Comparison of CNN architectures on the Microsoft Malware dataset.}
    
    \label{tab:microsoft_resnet_comparison}
    \centering
    \begin{tabular}{lccccc}
    
    \hline 
    Model & F1 Score & Precision & Recall \\
    
    \hline
    ResNet18 & .976 & .971 & .982 \\
    ResNet50 & .972 & .968 & .977 \\
    DenseNet121 & .978 & .976 & .980 \\
    Densenet169 & .981 & .983 & .979 \\
    MobileNetV2(x.5) & .956 & .960 & .953 \\
    MobileNetV2(x1) & .958 & .946 & .982 \\
    
    \midrule
    
    
    ResNet1D18 & .992 & .993 & .990 \\
    ResNet1D50 & .991 & .990 & .991 \\
    DenseNet1D121 & .980 & .971 & .992 \\
    Densenet1D169 & .985 & .980 & .991 \\
    MobileNet1DV2(x.5) & .984 & .985 & .982 \\
    MobileNet1DV2(x1) & .989 & .975 & .988 \\
    
    \hline
    \end{tabular}
    \end{table}

\subsection{Adapting Convolutions to 1D Data}
\label{sec:conv_adaptation_ablation}

This section investigates methodologies for converting 2D CNNs to 1D models. Two primary techniques were employed: the first preserves the original stride and kernel sizes, while the second squares these values to account for the dimensionality reduction when transitioning to 1D data. Alternatively, either the kernel size or the stride can be squared independently, maintaining the original size of the other parameter.

When the kernel size is squared, this is equivalent to flattening the 2D kernel into a 1D form, thereby preserving the same number of parameters. However, this transformation also extends the kernel's receptive field, making it sensitive to longer-range dependencies along the single dimension. Squaring the stride preserves the overall amount of downsampling after each ResNet stage block, thereby maintaining consistent VRAM requirements and computational load per forward pass. In contrast, preserving the original stride reduces the amount of downsampling, consequently increasing the VRAM and computational demands of the model.

In Table~\ref{tab:kernel_conversion_ablation} the model performance for each combination of 2D to 1D conversion methodology is shown for a ResNet18 model trained on the MalNet type classification dataset. Best performance was achieved when both the kernel sizes and stride values were squared, this also produced a model with the same number of parameters and compute requirements as the original 2D model. Flattening the kernel sizes appeared to have more of an impact on model performance than squaring the stride values. Additionally, squaring the stride improved performance while maintaining the same degree of downsampling as 2D models, meaning that training the model required the same amount of VRAM and compute per forward pass.

\begin{table*}[t]
  \centering
  
  \caption{Comparison of various approaches to convert 2D CNNs to 1D when training a ResNet18 on MalNet type classification.}
  \label{tab:kernel_conversion_ablation}
  \resizebox{\linewidth}{!}{
  \begin{tabular}{ccccccc}
  
  \hline 
  Squared Stride & Squared Kernel & F1 Score & Precision & Recall & Parameters (M) & Compute (GFLOPs)\\
  \hline
  
  False & False & .406 & .598 & .406 & 3.9 & 11.1 \\
  False & True & .471 & .561 & .434 & 11.2 & 32.6 \\
  True & False & .466 & .573 & .420 & 3.9 & 0.8 \\
  True & True & .482 & .598 & .432 & 11.2 & 2.3  \\
  
  \hline
  \end{tabular}}
  \end{table*}

\subsection{Comparison to Existing Models}
\label{sec:benchmarking}

The performance of the proposed approach was evaluated using precision, recall, and macro-averaged F1 score across binary, type, and family level classification tasks on the MalNet dataset. A ResNet1DV2-152D-SE model, trained with the improved recipe, was compared to three popular convolutional architectures: ResNet, DenseNet, and EfficientNet, as reported in the MalNet paper~\citep{malnet_dataset}. Additionally, it was compared to the current SOTA model, SHERLOCK, which employs a ViT-B pretrained using masked autoencoding~\citep{sherlock}. Due to the recency of the MalNet dataset these baselines encompass all models currently benchmarked.

The performance comparison results, presented in Table~\ref{tab:baseline_comp}, demonstrate the strong performance of the proposed model. The results indicate that the proposed model outperforms all baselines across all three classification granularities in terms of F1 score, thus achieving SOTA performance. Furthermore, the proposed model surpasses all baseline models in terms of recall and precision for both type and family classification tasks. While the proposed approach is outperformed by SHERLOCK in precision for binary classification, this can be attributed to the precision-recall trade-off, where a model's recall can be sacrificed to improve its precision and vice versa. While SHERLOCK favours precision, ResNet18 favours recall, making it feasible to individually outperform each model but challenging to surpass both models on both metrics simultaneously. However, the overall performance, as quantified by the F1 score, indicates the proposed model's superior overall performance compared to the baselines in binary classification. Adjusting the model's decision boundaries to make decisions closer to the marginal label distribution could potentially enhance its precision at the expense of recall.

\begin{table*}[t]
  \centering
  \caption{Performance comparison of the proposed approach with existing approaches on the MalNet dataset across task granularities.}
  \label{tab:baseline_comp}
  \resizebox{\linewidth}{!}{\begin{tabular}{lccccccccc}
      \toprule
      \multirow{2}{*}{Model} & \multicolumn{3}{c}{Binary} & \multicolumn{3}{c}{Type} & \multicolumn{3}{c}{Family} \\

       & F1 Score & Precision & Recall & F1 Score & Precision & Recall & F1 Score & Precision & Recall\\

      \midrule
      \textbf{ResNet1DV2-152D-SE}&
      \textbf{.874} & .907 & \textbf{.846} &
      \textbf{.503} & \textbf{.643} & \textbf{.453} &
      \textbf{.507} &  \textbf{.580} & \textbf{.480} 
      \\
      
      SHERLOCK & 
      .854 & \textbf{.920} & .810 & 
      .497 & .628 & .447 & 
      .491 & .568 & .461 
      \\ 
      
      ResNet18 & 
      .862 & .893 & .837 & 
      .467 & .556 & .424 & 
      .454 & .538 & .423 
      \\ 
      
      ResNet50 & 
      .854 & .907 & .814 & 
      .479 & .566 & .441 & 
      .468 & .541 & .443 
      \\ 
      
      DenseNet121 & 
      .864 & .900 & .834 & 
      .471 & .558 & .428 & 
      .461 & .529 & .438 
      \\ 
      
      Densenet169 & 
      .864 & .890 & .841
      & .477 & .573 & .433 & 
      .462 & .545 & .434 
      \\ 
      
      MobileNetV2(x.5) & 
      .857 & .894 & .827 & 
      .460 & .547 & .424 & 
      .451 & .528 & .423 
      \\ 
      
      MobileNetV2(x1) & 
      .854 & .889 & .825 & 
      .452 & .527 & .419 & 
      .438 & .532 & .405 
      \\
      
      \bottomrule
      
  \end{tabular}}
\end{table*}


\section{Additional Results and Ablations}
\label{sec:ablations}

The proposed ResNet1DV2-152D-SE model was shown to achieve state-of-the-art performance on the MalNet dataset. This section provides additional analysis and examines the contribution of various design decisions to the proposed model's performance. Section~\ref{sec:block_ablation} begins by comparing the performance of various residual block architectures. Next, architectural design choices are evaluated with various ResNet stems and activation functions being compared in Section~\ref{sec:stem_ablation} and Section~\ref{sec:activation_ablation}, respectively. The input representation is analysed in the following sections, with Section~\ref{sec:filter_ablation} examining the choice of resampling filter and Section~\ref{sec:signal_length_ablation} investigating the length of the input signals. Performance at different model sizes is compared in Section~\ref{sec:signal_length_ablation}. Finally, Section~\ref{sec:precision_recall_tradeoff} examines the trade-off between precision and recall by analysing the precision-recall (PR) curves of ResNet1DV2-152D-SE and SHERLOCK.

\subsection{Choice of ResNet Architecture}
\label{sec:block_ablation}

The performance of various residual block architectures is compared in Table~\ref{tab:resnet_block_ablation}. Specifically, the original ResNet block is evaluated alongside the bottleneck block, as well as the pre-activation block, both with and without squeeze-and-excitation modules, when incorporated into a ResNet1D18 model. While the ResNetV1.5 architecture achieved the best performance; adding SE layers to the ResNetV2 architecture resulted in an improvement in the model's F1-score and precision. This enhancement resulted in a marginal increase in computational overhead, but a large increase in model parameter count.

\begin{table*}[t]
    \caption{Performance comparison of different residual blocks when used in a ResNet1D18 trained on MalNet type classification.}
    \label{tab:resnet_block_ablation}
    \centering
    \resizebox{\linewidth}{!}{
    \begin{tabular}{lccccc}
    
    \hline 
    Model Architecture & F1 Score & Precision & Recall & Parameters (M) & Compute (GFLOPs)\\
    \hline    
    ResNetV1 & .518 & .645 & .464 & 11.2 & 2.3 \\
    ResNetV1.5 & .520 & .669 & .457 & 14.0 & 5.4\\
    ResNetV2 & .510 & .619 & .461 & 14.4 & 5.4 \\
    ResNetV2SE & .512 & .632 & .459 & 21.0 & 5.4\\
    
    \hline
    \end{tabular}}
    \end{table*}

\subsection{Effect of Model Stem}
\label{sec:stem_ablation}

The effect of replacing the initial convolution at the start of the ResNet with the deep stem containing multiple convolutions is shown in Table~\ref{tab:model_architecture}, where a ResNet1DV2-152-SE model was benchmarked on MalNet type classification with both a deep stem and with the original convolution. The deep stem improved the precision of the model, although this came at the expense of recall and F1 score. Additionally, the deep stem introduced a marginal increase in the number of parameters and a slight computational overhead.

\begin{table}[t]

    \caption{Performance comparison of stems when used in a ResNet1DV2-152-SE trained on MalNet type classification.}
    \label{tab:model_architecture}
    \centering
    \begin{tabular}{lcccccc}
    
    \hline 
    Stem Architecture & F1 Score & Precision & Recall & Parameters (M) & Compute (GFLOPs)\\
    \hline
    
    Standard & .506 & .591 & .465 & 106.8 & 29.4 \\
    Deep  & .503 & .643 & .453 & 106.8 & 29.8\\
    
    \hline
    \end{tabular}
    \end{table}

\subsection{Choice of Activation Function}
\label{sec:activation_ablation}

The proposed model replaces the ReLU function, which is typically used with ResNet architecture, with the GELU activation function. The impact of this on model performance is shown in Table~\ref{tab:activation_ablation}, where a ResNet1DV2-152D-SE model is trained using each activation function with the improved training recipe on MalNet type classification. The GELU activation function produced a high precision model, however, this came at the expense of recall, F1 score, and a marginal increase in computational overhead. The reason for this performance difference may be due to the GELU activation function learning smoother decision boundaries resulting in more reliable predictions. 

\begin{table*}[t]

    \caption{Performance comparison of ReLU and GELU activations when used in a ResNet1DV2-152D-SE for MalNet type classification.}
    \label{tab:activation_ablation}
    \centering
    \resizebox{\linewidth}{!}{
    \begin{tabular}{lccccc}
    
    \hline 
    Activation & F1 Score & Precision & Recall & Parameters (M) & Compute (GFLOPs) \\
    \hline
    ReLU & .509 & .580 & .473 & 106.8 & 29.8 \\
    GELU & .503 & .643 & .453 & 106.8 & 29.8 \\
    \hline
    \end{tabular}}
  \end{table*}

\subsection{Choice of Resampling Filter}
\label{sec:filter_ablation}
  
The Lanczos filter employed by this work has been widely adopted in prior works utilsing byteplot images; however, it was originally developed for natural images and thus may not be the ideal resampling filter for byte data. To evaluate its effectiveness, a ResNet1D18 model was trained MalNet type level classification using signal representations resampled using various filters. The results, given in Table~\ref{tab:resampling_ablation}, show that Lanczos resampling outperforms nearest neighbour, linear, and cubic resampling.

The superior performance of Lanczos resampling suggests that discriminative features in byteplot signals may reside in higher frequency components. Since the Lanczos filter preserves more high-frequency content due to its sinc-based formulation, it is better able to retain fine-grained variations in byte transitions that may be informative for classification. This implies that abrupt changes in byte values, which manifest as high-frequency components in the signal domain, are likely important for distinguishing malware types.

\begin{table}[t]

  \caption{Performance comparison of Resnet1D18 models when trained on signals generated using various resizing filters.}
  \label{tab:resampling_ablation}
  \centering
  \begin{tabular}{lccc}
  
  \hline 
  Activation & F1 Score & Precision & Recall \\
  \hline
  Lanczos & .518 & .645 & .464 \\
  Nearest Neighbour & .450 & 618 & 391 \\
  Cubic & .449 & .623 & .387 \\
  Linear & .287 & .517 & .227 \\
  \hline
  \end{tabular}
\end{table}

\subsection{Choice of Signal Length}
\label{sec:signal_length_ablation}

In this work's main evaluation, signal representations were generated by resizing file binaries to a length of $65536$. This was chosen to allow for comparison with existing models trained on the MalNet dataset which consists of byteplot images resized to $256\times 256$ ($65536$) pixels. In Table~\ref{tab:signal_length_ablation} the performance of a ResNet1D18 model on MalNet type classification is shown when trained on malware signals resized to various lengths. It can be seen that the F1 score, precision, and recall monotonically increase with signal length. This trend can be attributed to reduced information loss in longer signals, which retain more of the original binary's structure. However, this benefit comes at the cost of increased storage requirements as well as higher computational demands during both training and inference.

\begin{table}[t]

  \caption{Performance comparison ResNet1D18 models when trained on signal representations of varying length on MalNet type classification.}
  \label{tab:signal_length_ablation}
  \centering
  \begin{tabular}{lcccc}
  
  \hline 
  Signal Length & F1 Score & Precision & Recall & Compute (GFLOPs) \\
  \hline
  256 & .018 & .017 & .021 & 0.01 \\
  1024 & .034 & .169 & .034 & 0.03 \\
  4096 & .106 & .308 & .087 & 0.1 \\
  16384 & .174 & .387 & .135 & 0.6 \\
  65536 & .518 & .645 & .464 & 2.3 \\
  \hline
  \end{tabular}
\end{table}

\subsection{Parameter Sensitivity Analysis}
\label{sec:parameter_ablation}

The impact of the number of parameters on model performance is shown in Table~\ref{tab:paramter_ablation}, where the performance of each size of ResNet model is evaluated on MalNet type classification. Interestingly, small models appear to outperform their larger counterparts. This is hypothesised to be due to resizing file binaries significantly reducing the information in the input signal. As a consequence, the model performance becomes heavily dependent on signal length, with more parameters only resulting in increased overfitting.

\begin{table*}[t]

  \caption{Performance comparison of varying sizes of ResNet1D when trained on MalNet type classification.}
  \label{tab:paramter_ablation}
  \centering
  \resizebox{\linewidth}{!}{
  \begin{tabular}{lccccc}
  
  \hline 
  Model &  Parameters (M) & Compute (GFLOPs) & F1 Score & Precision & Recall \\
  \hline
  ResNet1D18 & 11.2 & 2.3  & .518 & .645 & .464 \\
  ResNet1D34 & 21.3 & 4.7 & .520 & .643 & .463 \\
  ResNet1D50 & 23.6 & 9.9 & .508 & .635 & .452  \\
  ResNet1D101 & 42.6 & 19.6 & .504 & .655 & .444 \\
  ResNet1D152 & 58.2 & 29.3 & .501 & .629 & .450 \\  
  \hline
  \end{tabular}}
\end{table*}

\subsection{Precision Recall Trade-off}
\label{sec:precision_recall_tradeoff}
\begin{figure*}[t]
  \centering
  \resizebox{0.6\linewidth}{!}{%
    \begin{tikzpicture}
      \begin{axis}[
          xlabel={Recall},
          ylabel={Precision},
          xmin=0.5, xmax=1.0,
          ymin=0.93, ymax=1,
          xtick={0.0,0.1,0.2,0.3,0.4,0.5,0.6,0.7,0.8,0.9,1.0},
          legend pos=south west,
          axis lines=left,  
          axis line style={-},
          grid=none,  
          ymajorgrids=true,  
          major grid style={line width=.1pt, draw=gray!25},  
          legend style={
              draw=gray!50,  
              line width=.1pt,  
              font = \footnotesize
          }
      ]
      
      \addplot[
          color=orange,
          line width=1pt
          ]
          coordinates {            
            (0.5037027142918198,0.9966939272100164)(0.5101258146797101,0.9967251369008625)(0.5169574269529887,0.9967303999863756)(0.5234343087650777,0.9967341784138242)(0.5304780822567609,0.9967675402704511)(0.5370899549175727,0.9967790055005302)(0.5436142608694018,0.996788384516334)(0.5499326351404347,0.996799999846574)(0.556465733081409,0.9968276493876534)(0.5627037696123118,0.9968296779647737)(0.5689179888700556,0.9968435912814667)(0.5750074558134444,0.9968662770504622)(0.5812999200173209,0.9968734561630778)(0.5878354376339404,0.9968936752713161)(0.5944494143605308,0.9969032718018862)(0.6011319723597145,0.9969177962563703)(0.6074468223205685,0.996917680370674)(0.6138095660787023,0.9969416400574603)(0.6205836171239018,0.9969506588049044)(0.626674857494161,0.9969658186533137)(0.6326077406137767,0.9969591063339807)(0.6386153030394827,0.9969585496441398)(0.6445908723935768,0.9969794933548444)(0.6505605353344502,0.9969979203010447)(0.6566781854874881,0.9969853297369078)(0.6627392174847866,0.9969743648301965)(0.669058861709807,0.996965640625493)(0.675380245546355,0.9969657474131107)(0.6814388090236242,0.9969677460737342)(0.6875294369604515,0.9969806087927869)(0.6934518824109024,0.99699258644024)(0.6993777958126274,0.9969875282083629)(0.7053415899128833,0.9969861562318619)(0.7111100703500028,0.9969850188326909)(0.7170012629091245,0.9969898283602088)(0.7229569338125719,0.9969897647662042)(0.7287507494560191,0.9969797489708423)(0.7349354892493042,0.9969734194510018)(0.741438063207451,0.9969724098926664)(0.747471813737577,0.9969700916999455)(0.7533251368699485,0.9969692643130075)(0.759184486647871,0.9969767169378)(0.7651120870594789,0.9969593165803631)(0.7709370090611029,0.9969394233673118)(0.7766928049873874,0.996910334811281)(0.7823233338546532,0.996899022797414)(0.7880477529000178,0.9969067568280845)(0.7942602513325359,0.996914808468704)(0.7999927635166265,0.9969072205402342)(0.8054753314725501,0.9968902665364288)(0.8109588387435533,0.9968867329069285)(0.8165276095229642,0.9968904490314692)(0.8219216226104412,0.9969011966102472)(0.827341305300414,0.9968889387758147)(0.8327386548350538,0.9968708378924976)(0.8380590030767265,0.9968562633279491)(0.8433395619316263,0.9968409825836295)(0.8486879393352749,0.9968173499698777)(0.8539775494302818,0.996785896060033)(0.8591948172351166,0.996763400122237)(0.8646096831173613,0.9967553363886641)(0.8699522856119003,0.9967441822549586)(0.8749982636760754,0.9967179497169507)(0.8800697347515112,0.9966813064058778)(0.8851917297064499,0.9966592783671986)(0.890276663045607,0.9966220309049625)(0.8953376100348909,0.9965977973775546)(0.900360349443996,0.9965435261061801)(0.9053646933065816,0.996502652482955)(0.910358355278082,0.9964718574414837)(0.9153779498603004,0.9964438191925054)(0.9203032785383881,0.9963892395764667)(0.925228866467438,0.9963428628013842)(0.9301408117842713,0.996292201845535)(0.9351159053752777,0.9962437012045977)(0.94011454947396,0.9961775622787411)(0.9450628274980732,0.996129314830857)(0.9499629073868208,0.9960623864337441)(0.9547898108736053,0.995983185644455)(0.9595936485392946,0.9958694265259069)(0.9644069883163295,0.995761828498681)(0.9691149519704716,0.9956558154805106)(0.9739377412572074,0.9955077856824945)(0.9790259809854449,0.9953059592078902)(0.9837934483712513,0.9950355856648258)(0.9892076229175972,0.9944193939500987)(0.9944582927666633,0.9933413388407895)(0.9976507504423001,0.991138772713605)(0.9988348623034542,0.9875142221334042)(0.9991495403696486,0.9831674112632871)(0.9992782904089824,0.9786608841830463)(0.9993554382351021,0.9741464396663476)(0.9993915981084073,0.9696038345533134)(0.9994627080171951,0.964994203270951)(0.9995007878505229,0.9605405435535739)(0.9995471073557294,0.9561698545933319)(0.9995899701814427,0.9518466802276266)(0.9996505297032561,0.9475804854850647)(0.9997648781637892,0.9434240288064284)(0.9999013230722553,0.9393234708700307)
          };
      \addlegendentry{ResNet1DV2-152D-SE (AUC = .4983)}
      
      \addplot[
        color=blue,
        line width=1pt
        ]
        coordinates {
          (0.5033127671667913,0.99934118896639)(0.5095208582232547,0.9993305969105102)(0.5165219814623975,0.9993195224713055)(0.5232488829769633,0.9993036465711332)(0.5292741058425785,0.9992667508376691)(0.5353277412752738,0.9992535833780151)(0.5415706349838844,0.9992473001296498)(0.5479106689556198,0.9992372172443803)(0.5542560919905127,0.9992375073528728)(0.5604299426126137,0.9992232612693043)(0.5664201502803339,0.9991778502564367)(0.5727326835748112,0.9991646616836185)(0.5796744619842338,0.9991467796491869)(0.5854075338629857,0.9991179283112822)(0.591125809815155,0.999071715463822)(0.5968868623131764,0.9990253158018597)(0.6027894051443414,0.9989967979563251)(0.6086442211706213,0.9989790445143721)(0.6142712978561364,0.9989635538905548)(0.6199223887830121,0.9989201874280601)(0.6257144969573339,0.998860351358139)(0.6315969610716327,0.998817010404537)(0.6373944000304392,0.9987651647379033)(0.643040509844867,0.9987390949944979)(0.6486726499186033,0.9986784406490324)(0.6543642959454496,0.9986196710982419)(0.6601310117592066,0.9985884068808756)(0.6658318143940005,0.9985219359325994)(0.6713964849053935,0.9984754045793516)(0.6770631396660701,0.9984461036186333)(0.6826939256177026,0.9984199871727667)(0.6883968948478626,0.9983663036229499)(0.6940743688727411,0.9983022079322322)(0.6999661747154775,0.9982314115089855)(0.7055308623676567,0.9981762267093176)(0.7110516010442126,0.9981088834410747)(0.7165864054498622,0.9980347571570976)(0.7219812964746919,0.9979393045981932)(0.7273998280717445,0.9978796212235119)(0.7329198639760692,0.9977950837515996)(0.7383963759962667,0.9977153509936663)(0.7438048630926504,0.9976233885167808)(0.7494581651809358,0.997563131031418)(0.7557354501628267,0.997490649377621)(0.7613876278155426,0.9974310238489518)(0.7668170506680987,0.9973548404502772)(0.7722958732662656,0.9972857165392573)(0.7781048959683352,0.9972301493937634)(0.7835176785457217,0.9971462173391713)(0.7889429190464634,0.9970336608040552)(0.7948255682812523,0.9969283125733887)(0.8005252087707468,0.9968181905623276)(0.8058208992099596,0.9966901445509312)(0.8111552592626692,0.9965923320358124)(0.8165037741765604,0.9964395020566782)(0.821961846139045,0.9962736241217579)(0.8272825758386283,0.9961192510076532)(0.8326697157999723,0.995953642803064)(0.8380263657309583,0.9957574563671)(0.8433491043306756,0.9955751014793701)(0.8488151844684376,0.9954464916000808)(0.8542132602513289,0.9952934354922225)(0.8594831743763679,0.9951307136072495)(0.8647228658673115,0.9949515275851935)(0.8699708295016391,0.9947154437052146)(0.875104474376988,0.9944727005635591)(0.8802835046258372,0.9942410702933284)(0.88556607550735,0.9940403647147771)(0.8907618488760857,0.9938066939756925)(0.8959015890149803,0.9935413622356434)(0.9009109871880269,0.9932431953114413)(0.9058677117253128,0.992885553155762)(0.9108712168960896,0.9925125362497089)(0.9164141426062054,0.9920578915838962)(0.9222718177156106,0.9916352282656079)(0.927357214706176,0.9912120039716878)(0.9324258171544967,0.9907635346689412)(0.9375769867695791,0.9902954632337398)(0.9425574444393332,0.9897475040012719)(0.947230087871863,0.989146656789898)(0.9518922719966997,0.9885356765381985)(0.9565569106821095,0.9878643903794642)(0.9610731855791399,0.9870849392682908)(0.9654898553680007,0.9860975334947888)(0.9697771773566087,0.9850541844079779)(0.9738923132922817,0.9839805614322538)(0.9779018414164841,0.9827848956915172)(0.9818782638263658,0.9814166320270159)(0.985666963774769,0.9799152617722939)(0.9892012670318264,0.9782523081961867)(0.9922996869540521,0.9762351274133966)(0.9949918085382967,0.9737716250568466)(0.9971809646071322,0.9708889052361911)(0.9988397436153886,0.967556340370464)(0.9997315067261456,0.9634926842093711)(0.9999926603153815,0.9587955521171746)(1.0,0.9539962765349943)(1.0,0.9490744967895709)(1.0,0.9442889471643781)(1.0,0.9396228747031364)
        };
      \addlegendentry{SHERLOCK (AUC = .0.4977)}
        
      \end{axis}
    \end{tikzpicture}
  }
  \caption{Precision-recall curves of ResNet1DV2-152D-SE and SHERLOCK on MalNet binary classification.}
  \label{fig:precision_recall_curves}
\end{figure*}
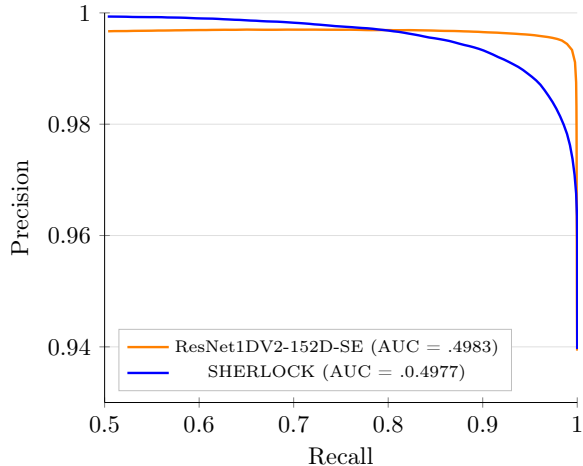

In Section~\ref{sec:benchmarking}, it was noted that while ResNet1DV2-152D-SE outperforms SHERLOCK in terms of recall, it exhibits lower precision. This was attributed to the trade-off between precision and recall, whereby altering a model's decision threshold can improve precision at the expense of recall and vice-versa. This trade-off can be analysed using the  PR curve, which visualises model performance across different decision thresholds. To focus on practical deployment scenarios---where maintaining a reasonable level of threat coverage is essential---the analysis was restricted to the region where recall is greater than or equal to $0.5$. The PR curves of the proposed model and SHERLOCK are shown in Figure~\ref{fig:precision_recall_curves}.

It can be seen that ResNet1DV2-152D-SE has a greater area under the PR curve than SHERLOCK indicating superior overall performance across varying decision thresholds. In particular, the decision threshold of ResNet1DV2-152D-SE can be adjusted to give a recall value greater than $0.9$ without a significant decrease in precision, whereas operating SHERLOCK in this region would result in significant precision degradation.

In practical scenarios, achieving high recall ensures that more malware samples are successfully detected, reducing the likelihood of missed threats. However, this often comes at the cost of lower precision, leading to a higher false positive rate and less reliable predictions. To balance these competing objectives, the model's decision threshold would typically be tuned on a held-out validation set to maximise recall while maintaining an acceptable false positive rate. This approach ensures the model remains effective in detecting malware without overwhelming analysts with excessive false alarms.

\section{Discussion and Conclusions}
\label{sec:conclusions}
Previous applications of deep neural networks in malware classification overwhelmingly preprocess the raw binaries into image representations known as byteplots. Computer vision approaches such as CNNs and vision transformers trained on these byteplots have shown competitive performance on malware classification tasks while also being resilient to obfuscation techniques such as packing and polymorphism. However, the byteplot representations used to train these models were initially designed to assist human analysts in quickly identifying regions of bytecode of interest for manual inspection and are not the ideal representation for training machine learning models. This work instead proposes a change in paradigm where malware binaries are resized in 1D into signal representations instead of images. These signal representations do not require heuristic reshaping rules to convert them into a 2D structure, furthermore, once resized they do not require quantisation into integer pixel values like images. 

This work quantitatively shows that by excluding the heuristic reshaping and quantisation steps, the proposed signal representations contain more information than their image-based counterparts. Furthermore, by converting 2D computer vision models to operate on 1D data, an equivalent model can be trained which has the same number of parameters and compute requirement as its 2D counterpart but boasts superior performance. This is shown to be true for both binaries obtained from android DEX files and from Windows' EXE files through a series of experiments on the MalNet dataset and the Microsoft Malware classification dataset.

To demonstrate the proposed signal representation's effectiveness, a novel 1D CNN architecture was proposed. The CNN was based on the ResNetV2 architecture, however, the 2D convolutions were replaced with 1D convolutions by the squaring the kernel sizes and stride values. Additionally, squeeze-and-excitation blocks were added to the model and the ReLU activation function was replaced with GELU. This model was compared to state-of-the-art approaches in binary, type and family level classification on the large-scale MalNet dataset. The proposed model achieved state-of-the-art performance with F1 scores of 0.974, 0.503, and 0.503 on binary, type, and family level classification respectively. 

These results pave the way for future works to use signal-based models instead of the current image-based models. By providing a method to adapt existing 2D CNNs to operate on 1D data it is hoped that researchers favour using 1D equivalent models for improved performance. In the future domain specific resampling methods could be used to generate the signal representations instead of Lanczos resampling, which was primarily intended for images. By preserving more relevant information from the original binaries in the signal representations the performance of the 1D models could be further improved. Additionally, it is hoped that future datasets will provide the raw binaries of samples. This would encourage research into alternative file representations, and allow the proposed approach to be evaluated on a greater variety of malware: for example cross-platform and mixed-format malware.

\section*{Declarations}

\subsection*{Conflicts of Interest/Competing Interests}
The authors declare that they have no competing interests.

\subsection*{Availability of Data and Materials}
The datasets used for this research are publically available.


\begin{thebibliography}{0}
\providecommand{\natexlab}[1]{#1}
\providecommand{\url}[1]{{#1}}
\providecommand{\urlprefix}{URL }
\providecommand{\doi}[1]{\url{https://doi.org/#1}}
\providecommand{\eprint}[2][]{\url{#2}}
 \bibcommenthead

\end{thebibliography}


\begin{thebibliography}{51}
  \providecommand{\natexlab}[1]{#1}
  \providecommand{\url}[1]{{#1}}
  \providecommand{\urlprefix}{URL }
  \providecommand{\doi}[1]{\url{https://doi.org/#1}}
  \providecommand{\eprint}[2][]{\url{#2}}
   \bibcommenthead
  
  \bibitem[{A. et~al.(2023)A., P., Yerima, Bashar, David, T., Antony, Shavanas, and T.}]{10040557}
  A. DK, P. V, Yerima SY, et~al (2023) Obfuscated malware detection in iot android applications using markov images and cnn. IEEE Systems Journal 17(2):2756--2766. \doi{10.1109/JSYST.2023.3238678}
  
  \bibitem[{Abusitta et~al.(2021)Abusitta, Li, and Fung}]{ABUSITTA2021102828}
  Abusitta A, Li MQ, Fung BC (2021) Malware classification and composition analysis: A survey of recent developments. Journal of Information Security and Applications 59:102828. \doi{https://doi.org/10.1016/j.jisa.2021.102828}, \urlprefix\url{https://www.sciencedirect.com/science/article/pii/S2214212621000648}
  
  \bibitem[{Allix et~al.(2016)Allix, Bissyand{\'e}, Klein, and Le~Traon}]{Allix:2016:ACM:2901739.2903508}
  Allix K, Bissyand{\'e} TF, Klein J, et~al (2016) Androzoo: Collecting millions of android apps for the research community. In: Proceedings of the 13th International Conference on Mining Software Repositories. ACM, New York, NY, USA, MSR '16, pp 468--471, \doi{10.1145/2901739.2903508}, \urlprefix\url{http://doi.acm.org/10.1145/2901739.2903508}
  
  \bibitem[{Alsulami and Mancoridis(2018)}]{8659358}
  Alsulami B, Mancoridis S (2018) Behavioral malware classification using convolutional recurrent neural networks. In: 2018 13th International Conference on Malicious and Unwanted Software (MALWARE), pp 103--111, \doi{10.1109/MALWARE.2018.8659358}
  
  \bibitem[{Anderson et~al.(2012)Anderson, Storlie, and Lane}]{10.1145/2381896.2381900}
  Anderson B, Storlie C, Lane T (2012) Improving malware classification: bridging the static/dynamic gap. In: Proceedings of the 5th ACM Workshop on Security and Artificial Intelligence. Association for Computing Machinery, New York, NY, USA, AISec '12, p 3–14, \doi{10.1145/2381896.2381900}, \urlprefix\url{https://doi.org/10.1145/2381896.2381900}
  
  \bibitem[{{Anderson} and {Roth}(2018)}]{2018arXiv180404637A}
  {Anderson} HS, {Roth} P (2018) {EMBER: An Open Dataset for Training Static PE Malware Machine Learning Models}. ArXiv e-prints {\href{https://arxiv.org/abs/1804.04637}{{arXiv:1804.04637}}} {[cs.CR]}
  
  \bibitem[{Bovenzi et~al.(2022)Bovenzi, Cerasuolo, Montieri, Nascita, Persico, and Pescapé}]{9912986}
  Bovenzi G, Cerasuolo F, Montieri A, et~al (2022) A comparison of machine and deep learning models for detection and classification of android malware traffic. In: 2022 IEEE Symposium on Computers and Communications (ISCC), pp 1--6, \doi{10.1109/ISCC55528.2022.9912986}
  
  \bibitem[{Chong et~al.(2022)Chong, Gao, Zhang, Liu, Huang, and Zhao}]{chong2022classification}
  Chong X, Gao Y, Zhang R, et~al (2022) Classification of malware families based on efficient-net and 1d-cnn fusion. Electronics 11(19):3064
  
  \bibitem[{Conti et~al.(2008)Conti, Dean, Sinda, and Sangster}]{10.1007/978-3-540-85933-8_1}
  Conti G, Dean E, Sinda M, et~al (2008) Visual reverse engineering of binary and data files. In: Goodall JR, Conti G, Ma KL (eds) Visualization for Computer Security. Springer Berlin Heidelberg, Berlin, Heidelberg, pp 1--17
  
  \bibitem[{Conti et~al.(2010)Conti, Bratus, Shubina, Sangster, Ragsdale, Supan, Lichtenberg, and Perez-Alemany}]{CONTI2010S3}
  Conti G, Bratus S, Shubina A, et~al (2010) Automated mapping of large binary objects using primitive fragment type classification. Digital Investigation 7:S3--S12. \doi{https://doi.org/10.1016/j.diin.2010.05.002}, \urlprefix\url{https://www.sciencedirect.com/science/article/pii/S1742287610000290}, the Proceedings of the Tenth Annual DFRWS Conference
  
  \bibitem[{Daly et~al.(2023)Daly, Fieldsend, Hassall, and Tabor}]{10.1088/2632-2153/aced7f}
  Daly G, Fieldsend J, Hassall G, et~al (2023) Data-driven plasma modelling: Surrogate collisional radiative models of fluorocarbon plasmas from deep generative autoencoders. Machine Learning: Science and Technology 4. \doi{10.1088/2632-2153/aced7f}
  
  \bibitem[{Damasevicius et~al.(2021)Damasevicius, Venčkauskas, Toldinas, and Grigaliunas}]{electronics10040485}
  Damasevicius R, Venčkauskas A, Toldinas J, et~al (2021) Ensemble-based classification using neural networks and machine learning models for windows pe malware detection. Electronics 10(4). \doi{10.3390/electronics10040485}, \urlprefix\url{https://www.mdpi.com/2079-9292/10/4/485}
  
  \bibitem[{Deng and Mirkovic(2022)}]{10.1109/COMSNETS53615.2022.9668396}
  Deng X, Mirkovic J (2022) Polymorphic malware behavior through network trace analysis. pp 138--146, \doi{10.1109/COMSNETS53615.2022.9668396}
  
  \bibitem[{Dib et~al.(2021)Dib, Torabi, Bou-Harb, and Assi}]{9411822}
  Dib M, Torabi S, Bou-Harb E, et~al (2021) A multi-dimensional deep learning framework for iot malware classification and family attribution. IEEE Transactions on Network and Service Management 18(2):1165--1177. \doi{10.1109/TNSM.2021.3075315}
  
  \bibitem[{Freitas et~al.(2021)Freitas, Dong, Neil, and Chau}]{malnet_dataset}
  Freitas S, Dong Y, Neil J, et~al (2021) A large-scale database for graph representation learning. {\href{https://arxiv.org/abs/2011.07682}{{arXiv:2011.07682}}}
  
  \bibitem[{Gennissen and Blasco(2017)}]{Gennissen2017GamutS}
  Gennissen J, Blasco J (2017) Gamut : Sifting through images to detect android malware. \urlprefix\url{https://api.semanticscholar.org/CorpusID:44430018}
  
  \bibitem[{Gibert et~al.(2020)Gibert, Mateu, and Planes}]{GIBERT2020101873}
  Gibert D, Mateu C, Planes J (2020) Hydra: A multimodal deep learning framework for malware classification. Computers \& Security 95:101873. \doi{https://doi.org/10.1016/j.cose.2020.101873}, \urlprefix\url{http://www.sciencedirect.com/science/article/pii/S0167404820301462}
  
  \bibitem[{Hasegawa and Iyatomi(2018)}]{8368693}
  Hasegawa C, Iyatomi H (2018) One-dimensional convolutional neural networks for android malware detection. In: 2018 IEEE 14th International Colloquium on Signal Processing \& Its Applications (CSPA), pp 99--102, \doi{10.1109/CSPA.2018.8368693}
  
  \bibitem[{He et~al.(2015)He, Zhang, Ren, and Sun}]{he2015deep}
  He K, Zhang X, Ren S, et~al (2015) Deep residual learning for image recognition. {\href{https://arxiv.org/abs/1512.03385}{{arXiv:1512.03385}}}
  
  \bibitem[{He et~al.(2016)He, Zhang, Ren, and Sun}]{he2016identity}
  He K, Zhang X, Ren S, et~al (2016) Identity mappings in deep residual networks. {\href{https://arxiv.org/abs/1603.05027}{{arXiv:1603.05027}}}
  
  \bibitem[{He et~al.(2018)He, Zhang, Zhang, Zhang, Xie, and Li}]{he2018bag}
  He T, Zhang Z, Zhang H, et~al (2018) Bag of tricks for image classification with convolutional neural networks. {\href{https://arxiv.org/abs/1812.01187}{{arXiv:1812.01187}}}
  
  \bibitem[{Hendrycks and Gimpel(2023)}]{hendrycks2023gaussian}
  Hendrycks D, Gimpel K (2023) Gaussian error linear units (gelus). {\href{https://arxiv.org/abs/1606.08415}{{arXiv:1606.08415}}}
  
  \bibitem[{Hu et~al.(2018)Hu, Shen, and Sun}]{squeeze_excite}
  Hu J, Shen L, Sun G (2018) Squeeze-and-excitation networks. In: Proceedings of the IEEE Conference on Computer Vision and Pattern Recognition (CVPR)
  
  \bibitem[{Huang and Kao(2018)}]{huang2018r2d2}
  Huang THD, Kao HY (2018) R2-d2: Color-inspired convolutional neural network (cnn)-based android malware detections. {\href{https://arxiv.org/abs/1705.04448}{{arXiv:1705.04448}}}
  
  \bibitem[{Kalash et~al.(2018)Kalash, Rochan, Mohammed, Bruce, Wang, and Iqbal}]{8328749}
  Kalash M, Rochan M, Mohammed N, et~al (2018) Malware classification with deep convolutional neural networks. In: 2018 9th IFIP International Conference on New Technologies, Mobility and Security (NTMS), pp 1--5, \doi{10.1109/NTMS.2018.8328749}
  
  \bibitem[{Kim et~al.(2022)Kim, Kang, Cho, and Choi}]{9665792}
  Kim HI, Kang M, Cho SJ, et~al (2022) Efficient deep learning network with multi-streams for android malware family classification. IEEE Access 10:5518--5532. \doi{10.1109/ACCESS.2021.3139334}
  
  \bibitem[{Kim et~al.(2023)Kim, Paik, and Cho}]{10061401}
  Kim J, Paik JY, Cho ES (2023) Attention-based cross-modal cnn using non-disassembled files for malware classification. IEEE Access 11:22889--22903. \doi{10.1109/ACCESS.2023.3253770}
  
  \bibitem[{Liu et~al.(2017)Liu, Wang, Yu, and Zhong}]{Liu2017}
  Liu L, Wang Bs, Yu B, et~al (2017) Automatic malware classification and new malware detection using machine learning. Frontiers of Information Technology \& Electronic Engineering 18(9):1336--1347. \doi{10.1631/FITEE.1601325}, \urlprefix\url{https://doi.org/10.1631/FITEE.1601325}
  
  \bibitem[{Llaurad{\'o}(2016)}]{Llaurad2016ConvolutionalNN}
  Llaurad{\'o} DG (2016) Convolutional neural networks for malware classification. \urlprefix\url{https://api.semanticscholar.org/CorpusID:22879106}
  
  \bibitem[{Lu et~al.(2022)Lu, Zhang, Kinawi, and Niu}]{9877977}
  Lu Q, Zhang H, Kinawi H, et~al (2022) Self-attentive models for real-time malware classification. IEEE Access 10:95970--95985. \doi{10.1109/ACCESS.2022.3202952}
  
  \bibitem[{McLaughlin et~al.(2017)McLaughlin, Martinez~del Rincon, Kang, Yerima, Miller, Sezer, Safaei, Trickel, Zhao, Doup\'{e}, and Joon~Ahn}]{10.1145/3029806.3029823}
  McLaughlin N, Martinez~del Rincon J, Kang B, et~al (2017) Deep android malware detection. In: Proceedings of the Seventh ACM on Conference on Data and Application Security and Privacy. Association for Computing Machinery, New York, NY, USA, CODASPY '17, p 301–308, \doi{10.1145/3029806.3029823}, \urlprefix\url{https://doi.org/10.1145/3029806.3029823}
  
  \bibitem[{Mekruksavanich and Jitpattanakul(2022)}]{10.3390/s22083094}
  Mekruksavanich S, Jitpattanakul A (2022) Deep residual network for smartwatch-based user identification through complex hand movements. Sensors 22:3094. \doi{10.3390/s22083094}
  
  \bibitem[{Mohaisen et~al.(2014)Mohaisen, West, Mankin, and Alrawi}]{6997496}
  Mohaisen A, West AG, Mankin A, et~al (2014) Chatter: Classifying malware families using system event ordering. In: 2014 IEEE Conference on Communications and Network Security, pp 283--291, \doi{10.1109/CNS.2014.6997496}
  
  \bibitem[{Nataraj et~al.(2011{\natexlab{a}})Nataraj, Karthikeyan, Jacob, and Manjunath}]{10.1145/2016904.2016908}
  Nataraj L, Karthikeyan S, Jacob G, et~al (2011{\natexlab{a}}) Malware images: visualization and automatic classification. In: Proceedings of the 8th International Symposium on Visualization for Cyber Security. Association for Computing Machinery, New York, NY, USA, VizSec '11, \doi{10.1145/2016904.2016908}, \urlprefix\url{https://doi.org/10.1145/2016904.2016908}
  
  \bibitem[{Nataraj et~al.(2011{\natexlab{b}})Nataraj, Yegneswaran, Porras, and Zhang}]{10.1145/2046684.2046689}
  Nataraj L, Yegneswaran V, Porras P, et~al (2011{\natexlab{b}}) A comparative assessment of malware classification using binary texture analysis and dynamic analysis. In: Proceedings of the 4th ACM Workshop on Security and Artificial Intelligence. Association for Computing Machinery, New York, NY, USA, AISec '11, p 21–30, \doi{10.1145/2046684.2046689}, \urlprefix\url{https://doi.org/10.1145/2046684.2046689}
  
  \bibitem[{Noever and Noever(2021)}]{noever2021virusmnist}
  Noever D, Noever SEM (2021) Virus-mnist: A benchmark malware dataset. {\href{https://arxiv.org/abs/2103.00602}{{arXiv:2103.00602}}}
  
  \bibitem[{Paik et~al.(2022)Paik, Jin, and Cho}]{https://doi.org/10.1111/coin.12521}
  Paik JY, Jin R, Cho ES (2022) Malware classification using a byte-granularity feature based on structural entropy. Computational Intelligence 38(4):1536--1558. \doi{https://doi.org/10.1111/coin.12521}, \urlprefix\url{https://onlinelibrary.wiley.com/doi/abs/10.1111/coin.12521}, {\href{https://arxiv.org/abs/https://onlinelibrary.wiley.com/doi/pdf/10.1111/coin.12521}{{https://onlinelibrary.wiley.com/doi/pdf/10.1111/coin.12521}}}
  
  \bibitem[{Priya and Sathya~Sofia(2023)}]{10061076}
  Priya V, Sathya~Sofia A (2023) Review on malware classification and malware detection using transfer learning approach. In: 2023 5th International Conference on Smart Systems and Inventive Technology (ICSSIT), pp 1042--1049, \doi{10.1109/ICSSIT55814.2023.10061076}
  
  \bibitem[{Rezende et~al.(2018)Rezende, Ruppert, de~Carvalho, The{\'o}philo, Ramos, and de~Geus}]{Rezende2018MaliciousSC}
  Rezende ERS, Ruppert GCS, de~Carvalho TJ, et~al (2018) Malicious software classification using vgg16 deep neural network’s bottleneck features. \urlprefix\url{https://api.semanticscholar.org/CorpusID:21388434}
  
  \bibitem[{Ronen et~al.(2018)Ronen, Radu, Feuerstein, Yom-Tov, and Ahmadi}]{microsoft_malware_dataset}
  Ronen R, Radu M, Feuerstein C, et~al (2018) Microsoft malware classification challenge. {\href{https://arxiv.org/abs/1802.10135}{{arXiv:1802.10135}}}
  
  \bibitem[{Safa et~al.(2019)Safa, Nassar, and Rahal Al~Orabi}]{8766515}
  Safa H, Nassar M, Rahal Al~Orabi WA (2019) Benchmarking convolutional and recurrent neural networks for malware classification. In: 2019 15th International Wireless Communications \& Mobile Computing Conference (IWCMC), pp 561--566, \doi{10.1109/IWCMC.2019.8766515}
  
  \bibitem[{Sathyanarayan et~al.(2008)Sathyanarayan, Kohli, and Bezawada}]{10.1007/978-3-540-70500-0_25}
  Sathyanarayan V, Kohli P, Bezawada B (2008) Signature generation and detection of malware families. pp 336--349, \doi{10.1007/978-3-540-70500-0_25}
  
  \bibitem[{Schofield(2021)}]{Schofield2021}
  Schofield M (2021) Comparison of malware classification methods using convolutional neural network based on api call stream. International Journal of Network Security \& Its Applications (IJNSA) 13(2). Available at SSRN: \url{https://ssrn.com/abstract=3822934}
  
  \bibitem[{Schofield et~al.(2021)Schofield, Alicioglu, Binaco, Turner, Thatcher, Lam, and Sun}]{10.5121/csit.2021.110106}
  Schofield M, Alicioglu G, Binaco R, et~al (2021) Convolutional neural network for malware classification based on api call sequence. pp 85--98, \doi{10.5121/csit.2021.110106}
  
  \bibitem[{Schultz et~al.(2001)Schultz, Eskin, Zadok, and Stolfo}]{Schultz200138}
  Schultz MG, Eskin E, Zadok E, et~al (2001) Data mining methods for detection of new malicious executables. Proceedings of the IEEE Computer Society Symposium on Research in Security and Privacy p 38 – 49. \urlprefix\url{https://www.scopus.com/inward/record.uri?eid=2-s2.0-0034838197&partnerID=40&md5=388ce70cc320d3c766cada92e1b12ccf}, cited by: 840
  
  \bibitem[{Seneviratne et~al.(2022{\natexlab{a}})Seneviratne, Shariffdeen, Rasnayaka, and Kasthuriarachchi}]{Seneviratne_2022}
  Seneviratne S, Shariffdeen R, Rasnayaka S, et~al (2022{\natexlab{a}}) Self-supervised vision transformers for malware detection. IEEE Access 10:103121–103135. \doi{10.1109/access.2022.3206445}, \urlprefix\url{http://dx.doi.org/10.1109/ACCESS.2022.3206445}
  
  \bibitem[{Seneviratne et~al.(2022{\natexlab{b}})Seneviratne, Shariffdeen, Rasnayaka, and Kasthuriarachchi}]{sherlock}
  Seneviratne S, Shariffdeen R, Rasnayaka S, et~al (2022{\natexlab{b}}) Self-supervised vision transformers for malware detection. IEEE Access 10:103121–103135. \doi{10.1109/access.2022.3206445}, \urlprefix\url{http://dx.doi.org/10.1109/ACCESS.2022.3206445}
  
  \bibitem[{Votipka et~al.(2020)Votipka, Rabin, Micinski, Foster, and Mazurek}]{247696}
  Votipka D, Rabin S, Micinski K, et~al (2020) An observational investigation of reverse {Engineers{\textquoteright}} processes. In: 29th USENIX Security Symposium (USENIX Security 20). USENIX Association, pp 1875--1892, \urlprefix\url{https://www.usenix.org/conference/usenixsecurity20/presentation/votipka-observational}
  
  \bibitem[{Wang et~al.(2021)Wang, Sun, Luo, and Yang}]{10.32604/cmes.2022.018492}
  Wang L, Sun J, Luo X, et~al (2021) Transferable features from 1d-convolutional network for industrial malware classification. Computer Modeling in Engineering \& Sciences 130:1003--1016. \doi{10.32604/cmes.2022.018492}
  
  \bibitem[{Yeboah and Musah(2022)}]{paul2022nlp}
  Yeboah PN, Musah HBB (2022) Nlp technique for malware detection using 1d cnn fusion model. Security and Communication Networks 2022:2957203. \doi{10.1155/2022/2957203}, \urlprefix\url{https://doi.org/10.1155/2022/2957203}
  
  \bibitem[{Zyout et~al.(2023)Zyout, Shatnawi, and Najadat}]{Zyout2023}
  Zyout M, Shatnawi R, Najadat H (2023) Malware classification approaches utilizing binary and text encoding of permissions. International Journal of Information Security 22(6):1687--1712. \doi{10.1007/s10207-023-00712-z}, \urlprefix\url{https://doi.org/10.1007/s10207-023-00712-z}
  
  \end{thebibliography}
\end{document}